%% file: main_arXiv1.tex
\documentclass[10pt,twocolumn,twoside]{IEEEtran}
\IEEEoverridecommandlockouts
\usepackage{cite}
\usepackage{amsmath,amssymb,amsfonts,amsthm,commath}
\usepackage{graphicx}
\usepackage{textcomp}
\def\BibTeX{{\rm B\kern-.05em{\sc i\kern-.025em b}\kern-.08em
    T\kern-.1667em\lower.7ex\hbox{E}\kern-.125emX}}

\usepackage{pifont}
\usepackage{multicol}
\usepackage{lettrine}
\usepackage{comment}
\usepackage{enumitem}
\usepackage{booktabs}
\usepackage[table]{xcolor}
\usepackage{bbm}
\usepackage{bm}
\usepackage{chngpage}
\usepackage{textcomp}
\usepackage{hyperref}
\usepackage{url}
\usepackage{orcidlink}
\usepackage[normalem]{ulem}
\usepackage{algpseudocode}
\newsavebox{\ieeealgbox}

\usepackage{array}

\newtheorem{theorem}{Theorem}
\newtheorem*{theorem*}{Theorem}
\newtheorem{proposition}{Proposition}
\newtheorem{corollary}{Corollary}
\newtheorem{lemma}{Lemma}
\newtheorem{definition}{Definition}

\newtheorem{axiom}{Axiom}
\newtheorem*{policy1*}{OEs-aware pricing policy (Dynamic price $\pi^\ast$)}
\newtheorem*{policy2*}{OEs-aware pricing policy (Non-volumetric charge $A^\ast$)}
\newtheorem*{policyy*}{Member-level-OEs-aware pricing policy (D-NEM)}

\newcommand*{\mcD}{\mathcal D}

 \def\old#1{}

\input LTmacros

\begin{document}

%\title{A Decentralized Pricing Mechanism for Energy Communities under Operating Envelopes
%}
%\title{Dynamic Two-Part Pricing for Efficient Energy Communities under Operating Envelopes
\title{A Decentralized Market Mechanism for Energy Communities under Operating Envelopes
}

\author{\vspace{-0.2cm}\IEEEauthorblockN{Ahmed S. Alahmed\orcidlink{0000-0002-4715-4379}, Guido Cavraro\orcidlink{0000-0003-0296-720X}, Andrey Bernstein\orcidlink{0000-0003-4489-8388}, and Lang Tong\orcidlink{0000-0003-3322-2681}\thanks{\scriptsize Ahmed S. Alahmed and Lang Tong are with the School of Electrical and Computer Engineering
\textit{Cornell University}, Ithaca, NY, USA ({\tt \{\tcb{ASA278,~LT35}\}\tcb{@cornell.edu}}).} \thanks{\scriptsize Guido Cavraro and Andrey Bernstein are with the Power System Engineering Center, National Renewable Energy Laboratory, Golden, CO, USA ({\tt \{\tcb{Guido.Cavraro,~Andrey.Bernstein}\}\tcb{@nrel.gov}}).}
%\thanks{\scriptsize Mathematical proofs and additional theoretical results are available in \cite{Alahmed&Cavraro&Bernstein&Tong:24ExtensionArXiv}.}
}}

\maketitle

\begin{abstract}
We propose an operating envelopes (OEs) aware energy community market mechanism that dynamically charges/rewards its members based on two-part pricing. The OEs are imposed exogenously by a regulated distribution system operator (DSO) on the energy community's revenue meter and is subject to a generalized net energy metering (NEM) tariff design. By formulating the interaction of the community operator and its members as a Stackelberg game, we show that the proposed two-part pricing achieves a Nash equilibrium and maximizes the community's social welfare in a decentralized fashion while ensuring that the community's operation abides by the OEs. The market mechanism conforms with the cost-causation principle and guarantees community members a surplus level no less than their maximum surplus when they autonomously face the DSO. The dynamic and uniform community price is a monotonically decreasing function of the community's aggregate renewable generation. We also analyze the impact of exogenous parameters such as NEM rates and OEs on the value of joining the community. Lastly, through numerical studies, we showcase the community's welfare, and pricing, and compare its members' surplus to customers under the DSO's regime.
\end{abstract}


\vspace{-0.28cm}
\section{Introduction and Background}\label{sec:intro}
\input{introG}

\section{OEs-aware Energy Sharing: Models, Pricing Axioms, and Stackelberg Game}\label{sec:model}
\input{ECmodel}

\section{Two-Part Pricing and Stackelberg Equilibrium}\label{sec:MktMech}
\input{MktMech}
\section{Value of Community}\label{sec:VoC}
\input{VoC}

\section{Numerical Study}\label{sec:num}
\input{num}

\section{Conclusion}\label{sec:conclusion}
This work proposes an energy community market mechanism that incorporates the DSO's dynamic OEs into its pricing structure. The mechanism induces a collective member response that satisfies the community's operational limits and decentrally maximizes the community's social welfare while meeting the pricing axioms for equitable welfare allocations. The market mechanism charges its members via a two-part dynamic and resource-aware pricing, whereby, the OEs-awareness and welfare optimality are achieved through the dynamic price, and the non-volumetric part of the pricing is designed to ensure profit-neutrality, and member rationality, characterized by achieving benefits higher than possible outside the community. The community price is a monotonically decreasing function of the aggregate renewable generation in the community. The two-part pricing is threshold-based, with thresholds that can be computed {\em apriori} and with minimal member information. We analyzed the value of joining the community for each member and established multiple properties, including net-consumption complementarity.

A major policy implication of this work is that the proposed mechanism competes with the lucrative legacy NEM program, creating a business case for energy communities to exist under NEM. Another implication is that community operators can internalize the dynamic OEs into their pricing without losing the intuitive structural properties of the price and the low computational complexity.

Several limitations of this work are worth pursuing in future directions. Firstly, we assumed rational prosumers who react to the price by maximizing their own surplus, which might not be the case in practice, as there exists significant empirical evidence showing that most prosumers exercise price-unaware consumption decisions. Second, in scenarios where members may cooperate, even partially, to achieve higher benefits, it would be imperative to model and analyze such strategic and price-affecting cooperation behaviors. Lastly, it would be interesting to extend this work to the scenario of curtailable renewable DG, which adds more complexity to prosumers' and operator's decisions.

% Uncomment for bounded rationality
\begin{comment}
    Lastly, we show that the market mechanism can be generalized to communities that include irrational members.
\end{comment}

\section{Acknowledgment}
The authors appreciate the valuable insights and discussions provided by Prof. Timothy D. Mount from Cornell University.

The work of Ahmed S. Alahmed and Lang Tong was supported in part by the National Science Foundation under Award 2218110 and the Power Systems and Engineering Research Center (PSERC) under Research Project M-46. This work was authored in part by the National Renewable Energy Laboratory, operated by Alliance for Sustainable Energy, LLC, for the U.S. Department of Energy (DOE) under Contract No. DE-AC36-08GO28308. Funding provided by the NREL Laboratory Directed Research and Development Program. The views expressed in the article do not necessarily represent the views of the DOE or the U.S. Government. The U.S. Government retains and the publisher, by accepting the article for publication, acknowledges that the U.S. Government retains a nonexclusive, paid-up, irrevocable, worldwide license to publish or reproduce the published form of this work, or allow others to do so, for U.S. Government purposes.

\appendix 
For brevity, the mathematical proofs and additional theoretical results are all based on having $K=1$ and $\underline{d}_i=0$ for every prosumer $i\in \Nc$.

\subsection*{Appendix A: Pricing Policy under Member-Level OEs}\label{sec:MemberLevelOEsPolicy}
 \input{appendixA/appendixA}

 %\begin{comment}
\subsection*{Appendix B: Benchmark Prosumer Optimal Decision}\label{sec:BenchmarkPolicy}
\input{appendixB_v2/BenchmarkLem_v2}
 %\end{comment}

% \section{Acknowledgment}
%The work of Ahmed S. Alahmed and Lang Tong was supported in part by the National Science Foundation under Award 2218110 and the Power Systems and Engineering Research Center (PSERC) under Research Project M-46. This work was authored in part by the National Renewable Energy Laboratory, operated by Alliance for Sustainable Energy, LLC, for the U.S. Department of Energy (DOE) under Contract No. DE-AC36-08GO28308. Funding is provided by the U.S. Department of Energy Office of Energy Efficiency and Renewable Energy Building Technologies Office, United States. The views expressed in the article do not necessarily represent the views of the DOE or the U.S. Government. The U.S. Government retains and the publisher, by accepting the article for publication, acknowledges that the U.S. Government retains a nonexclusive, paid-up, irrevocable, worldwide license to publish or reproduce the published form of this work, or allow others to do so, for U.S. Government purposes.

%\begin{comment}
%\newpage
\section*{Appendix C: Proofs and additional theoretical results}\label{sec:AppProofs}

\input{appendixB_v2/appendixB_v2}

%\end{comment}

{
\bibliographystyle{IEEEtran}
\bibliography{EC_TCNS}
}

\end{document}

%% file: LTmacros.tex
\def\nn{\nonumber}
\def\beq{\begin{equation}}
\def\eeq{\end{equation}}
\def\bea{\begin{eqnarray}}
\def\eea{\end{eqnarray}}
\def\ba{\begin{array}}
\def\ea{\end{array}}

\def\bitem{\begin{itemize}}
\def\eitem{\end{itemize}}
\def\ben{\begin{enumerate}}
\def\een{\end{enumerate}}

\def\ie{{\it i.e.,\ \/}}

\definecolor{bgrd}{rgb}{1,1,1}
\definecolor{gray}{rgb}{0.5,0.5,0.5}
\definecolor{dkr}{rgb}{0.7,0.1,0.2}
\definecolor{dkb}{rgb}{0.1,0.1,0.8}

\def\tcb{\textcolor{blue}}

\newcommand{\mbbE}{\mathbb{E}}

\def\Ac{{\cal A}}

\def\Kc{{\cal K}}
\def\Lc{{\cal L}}

\def\Nc{{\cal N}}

\def\Pc{{\cal P}}

\def\Tc{{\cal T}}

\def\Xc{{\cal X}}

\DeclareMathOperator{\sgn}{sgn}

%% file: introG.tex
Primarily driven by sustainability and economic incentives, low-voltage distribution networks have experienced a massive deployment of behind-the-meter (BTM) distributed energy resources (DER), such as rooftop solar, electric vehicles, and battery energy storage.
This introduces a paradigm shift in distribution networks, where the nodes can be referred to as prosumers, \ie nodes both able to produce and consume energy. 
The DERs’ uncoordinated power injections could pose challenges to system stability and power quality~\cite{Liu&Ochoa&Riaz&Mancarella&Ting&San&Theunissen:21PEM} given also that the DER outputs cannot be directly measured (because of {\em load masking}) or controlled (due to deregulated electricity markets unbundled model) by the DSO.
To ensure that the net power injections do not compromise the network's operation, the DSO announces dynamic OEs, \ie upper and lower bounds on the net power injections for each bus or prosumer node. OEs vary spatially and temporally depending on the network's conditions and provide higher flexibility than the {\em fixed export limits} (e.g., 5kW or 3.5kW \cite{Liu&Ochoa&Riaz&Mancarella&Ting&San&Theunissen:21PEM}), which, as the level of penetration increases, quickly become obsolete. A pioneer in the implementation of OEs due to massive rooftop solar installations, Australia has moved to mandate OEs-capable equipment for BTM generation installed after April 2024 \cite{AustraliaOEs:23}. Many other countries are expected to follow, as OEs are becoming an essential part of active distribution networks that suffer from reverse power flows \cite{Gustavo_ReversePowerFlows:20IEEETIP}.

Analogous to network constraints at the transmission level, OEs are essentially conservative inner approximations of the power injections feasible region where a solution for the power flow equations is guaranteed to exist and meet the operational constraints. Though inner hyper-rectangle approximations are widely adopted, e.g., see~\cite{liu2023robust}, more advanced solutions propose ellipsoidal inner approximations~\cite{Cui2021Network} or convex restrictions of the feasibility sets~\cite{Lee2019Convex,Lankeshwara2023TimeVarying}. Such approaches, however, were proposed at the substation level, as they are practically difficult to implement at each prosumer or aggregator node in the distribution network.

Rather than directly facing the DSO, prosumers can organize themself into {\em energy communities}, \ie a group of spatially co-located nodes that perform energy and monetary transactions as a single entity behind the DSO's revenue meter that measures the community's net consumption~\cite{Yang&Guoqiang&Spanos:21TSG,Ferro2022Architecture}.
Under the widely adopted NEM tariff design, when the community is net-importing, it faces a {\em buy (retail) rate}; whereas when it is net-exporting, it faces a {\em sell (export) rate}~\cite{NEMevolution:23NAS,Alahmed&Tong:22IEEETSG}. 
Energy communities also appeal to nodes without BTM DERs, that can own quotes of a shared central PV or take advantage of other members' generation capabilities.

This paper proposes a novel two-part pricing\footnote{A two-part pricing comprises of a {\em volumetric charge}, that depends on the prosumer's net consumption, and a {\em non-volumetric charge}, which is a lump sum that does not vary with prosumer’s usage \cite{WalterOi:71QJE}.} mechanism for energy communities. First, the pricing mechanism is threshold-based with the energy price being a function of the community's aggregate supply. Second, the mechanism is OEs-aware, as it induces a collective response that does not violate the OEs on the community's aggregate net consumption. Third, the mechanism aligns each member's incentive with that of the community such that each member maximizing its individual surplus results in maximum community welfare. Fourth, the market mechanism obeys the cost-causation principle which, in part, ensures that joining the community is advantageous over autonomously facing the DSO.  

Despite the abundant literature on energy communities and DER aggregation, the majority of literature that considered pricing-based energy management market mechanisms %and cost allocation rules
neglected network and grid constraints \cite{Yang&Guoqiang&Spanos:21TSG,Han&Morstyn&McCulloch:19TPS,Chen&Zhao&Low&Mei:21TSG,Chakraborty&Poolla&Varaiya:19TSG, Vespermann&Hamacher&Kazempour:21TPS}.%,Radoszynski&Dvorkin&Pinson}.  
 Recently, however, there has been a growing work on network-cognizant energy community market mechanisms that incorporate the distribution network's voltage and thermal constraints \cite{Chen&Zhao&Low&Wierman:23TSG,Mediwaththe&Blackhall:21TPS,Jhala&Natarajan&Pahwa:19TSG}. A major barrier facing such work is that it assumes that the DER aggregator or energy community has access to the distribution network data, which might be complex, especially when considering network multi-phases and losses. Another barrier is the significant computation burden due to coupling between the network's buses, which also creates consumption dependencies across users. For example, to reduce the voltage at prosumer $i$'s bus, prosumer $j$, whose voltage is within limits, must increase its consumption.

Another line of work internalizes network constraints through the {\em distribution locational marginal prices} (dLMP) \cite{Papavasiliou:18TSG,Bai&Wang&Chen&Li:18TPS}; an extension to LMP theory. dLMP has been proposed as an alternative to inefficient DSO tariffs that lump network costs into a volumetric charge that lacks temporal and spatial variations. Considerable regulatory barriers hurdle such a pricing mechanism, as it assumes that the DSO uses dLMP to influence prosumers' and aggregators' decisions.

OEs enable DSOs to ensure network integrity, \ie satisfying voltage and thermal limits, without the need to directly control BTM DER or to share full network information with aggregators and energy community operators. The DSO-imposed OEs are announced to the aggregators and operators, typically day-ahead, who in turn need to take them into account by either direct control or through a pricing mechanism. The literature on OEs largely ignored incorporating them into a price-driven mechanism design that induces community members to collectively react to ensure a safe community operation \cite{Liu&Ochoa&Wong&Theunissen:21TSG,Blackhall:20ARENA,Gerdroodbari&Khorasany&Razzaghi:22AE,Yi&Verbic:22EPSR}. Rather, it focuses on methods to compute OEs, e.g., see~\cite{ross2020method} and to allocate OEs, e.g., see~\cite{AlamEtal:24TSE}. 
Alternatively, works in the literature treat OEs as constraints that need to be ensured while controlling or optimizing the network performance. A model predictive control scheme for building management is devised in~\cite{gasser2021predictive}; whereas~\cite{ross2021strategies} proposes a method to control thermostatic loads to provide frequency regulation.
In \cite{Azimetal:23TSG}, the authors consider an energy community with operator-designed OEs to maximize energy transactions between its members without compromising network constraints and is perhaps the closest to our work. The authors adopt an \textit{ex-post} allocation rule (Shapley value) to distribute the coalition welfare, whereas, in our case, \textit{ex-ante} and resource-aware pricing and allocation mechanisms are designed to distribute the coalition welfare and incentivize joining the coalition.

Under the proposed OEs-aware pricing mechanism, the profit-neutral operator charges/compensates its members via a {\em two-part pricing}, consisting of a threshold-based dynamic price and a fixed reward both of which are functions of the {\em aggregate DER} in the community.
The dynamic, yet uniform, price is used to charge the {\em net-consumption} of members; the fixed reward is applied only when the community is congested, \ie when one of the OEs binds. Both the community price and fixed rewards are endogenously determined, based on community members' flexibility and aggregate DER. The price monotonically decreases as the community aggregate generation/load ratio increases, indicating that the excess generation from net-producing members is first pooled to net-consuming members before it is exported back to the grid. 
% Footnote
%This is analogous to uplifts in wholesale electricity markets.

The paper contributions with respect to our prior work on the {\em Dynamic NEM} (D-NEM) mechanism \cite{Alahmed&Tong:24TEMPR,Alahmed&Cavraro&Bernstein&Tong:23AllertonArXiv} are fourfold. First, we incorporate community-level-OEs, which give rise to a two-part pricing mechanism that invites generalizing the cost-causation principle. Second, unlike D-NEM, 
the community prices may not lie within the DSO NEM X rates. Third, the pricing mechanism obeys a {\em four-threshold policy} that takes into account the community's OEs. Fourth, this work analyzes the value of joining the community over facing the DSO in a standalone setting, in addition to comparing the surplus of members under different OEs arrangements.

Table \ref{tab:MajorSymbols} lists the major symbols used throughout the paper. The notations used here are standard.   When necessary,  boldface letters indicate column vectors as in $\bm{x}=(x_1,\ldots, x_n)$. In particular, $\bm{1}$ is a column vector of all ones. For a vector $\bm{x}$,  $\bm{x}^\top$ is the transpose of $\bm{x}$.  For a multivariate function $f$ of $\bm{x}$, we use interchangeably $f(\bm{x})$ and $f(x_1,\ldots,x_n)$. For vectors $\bm{x},\bm{y}$,  $\bm{x} \preceq \bm{y}$ is the element-wise inequality $x_i \le y_i$ for all $i$, and $[\bm{x}]^+, [\bm{x}]^-$ are the element-wise positive and negative parts of vector $\bm{x}$, \ie $[x_i]^+:=\max\{0,x_i\}$, $[x_i]^- :=-\min\{0,x_i\}$ for all $i$, and $\bm{x}= [\bm{x}]^+ - [\bm{x}]^-$. 
%For brevity, we use the notation
%$[x]_{\underline{x}}^{\overline{x}}:= \max\{\underline{x},\min\{x,\overline{x}\}\}$ 
Also, $[x]_{\underline{x}}^{\overline{x}}$ denotes the projection of $x$ into the closed and convex set $[\underline{x},\overline{x}]$ as per the rule $[x]_{\underline{x}}^{\overline{x}}:=\max\{\underline{x},\min\{x,\overline{x}\}\}$. The notation is also used for vectors, \ie $[\bm{x}]_{\underline{\bm{x}}}^{\overline{\bm{x}}}$.
%for scalars $x,\underline{x},\overline{x}$. 
Lastly, we denote by $\mathbb{R}_+\hspace{-0.13cm}:=\{ x\in \mathbb{R}\hspace{-0.05cm}: x \geq 0\}$ and $\mathbb{R}_-\hspace{-0.13cm}:=\{ x\in \mathbb{R}\hspace{-0.05cm}: x \leq 0\}$ the set of non-negative and non-positive real numbers, respectively.

\begin{table}
\centering
\caption{Major variables and parameters (alphabetically ordered).}
\label{tab:MajorSymbols}
\vspace{-0.2cm}
\resizebox{\columnwidth}{!}{%
\begin{tabular}{@{}ll@{}}
\midrule \midrule
Symbol                                                   & Description                                                            \\ \midrule
$A^\ast(\cdot)$                                      & Non-volumetric charge of the OEs-aware pricing policy.        \\
$\bm{d}_i, d_{\Nc}$                                      & Consumption of member $i$ and community.         \\
$\bm{d}_i^\psi$                                          & Consumption of member $i$ under consumption policy $\psi$.      \\
$\overline{\bm{d}}_i$              & Consumption bundle's upper limit of member $i$.             \\
$r_i, r_{\Nc}$                                           & Generation of member $i$ and community.           \\
$i,\Nc$                                                  & Index and set of community members.                                    \\
$k, \Kc$                                                 & Index and set of consumption devices.                                  \\
$\bm{L}_i(\cdot), \bm{f}_i(\cdot)$                                        & Marginal and inverse utility function of member $i$.                               \\
$P^\chi_i(\cdot), P^{\mbox{\tiny NEM}}_i(\cdot)$             & Payment function of prosumer $i$ under policy $\chi$ and under NEM.                    \\
$\tilde{P}^\chi_i(\cdot)$             & Fixed-charge-adjusted payment function of prosumer $i$ under policy $\chi$.                    \\
$\pi^+, \pi^-$                                           & NEM X buy (retail) and sell (export) rates.                            \\
$\pi^{\mbox{\tiny NEM}}(\cdot), \pi^\ast(\cdot)$      & NEM X price and community OEs-aware pricing policy.\\
$\psi_{i,\chi}$                              & Community members’ consumption policy under pricing policy $\chi$.\\
$S^\chi_i(\cdot), S^{\mbox{\tiny NEM}}_i(\cdot)$             & Surplus function of prosumer $i$ under policy $\chi$ and under NEM.                    \\
$\sigma_1, \sigma_2, \sigma_3, \sigma_4$        & Thresholds of the OEs-aware pricing policy. \\
$t,\Tc$                                                  & Index and set of time.                                                 \\
$U_i(\cdot)$                                             & Utility of consumption of member $i$.                                  \\
$\mbox{VoC}^{\chi^\ast,\mbox{\tiny NEM}}_i$         & Value gained by member $i$ after joining a community policy $\chi^\ast$.\\
$W$                                                      & Community social welfare.                                              \\
$z_i, z_{\Nc}$                                           & Net consumption of member $i$ and community. \\
$z_i^\psi$                                               & Net consumption of member $i$ under consumption policy $\psi$.  \\
$\underline{z}_i, \overline{z}_i$ ($\underline{z}_\Nc, \overline{z}_\Nc$)                      & Export and import OEs of prosumer $i$ (the community).             \\
\midrule \midrule
\end{tabular}%
}
\end{table}

In $\S$\ref{sec:model}, we model the OEs-constrained energy community and its DER in addition to delineating the community pricing axioms and the Stackelberg game. In $\S$\ref{sec:MktMech}, we present the OEs-aware pricing policy and Stackelberg equilibrium. In $\S$\ref{sec:VoC}, we define and analyze the value of joining the community. Lastly, we present a numerical study to showcase the community performance compared to the benchmark in $\S$\ref{sec:num}, followed by a summary of our findings in $\S$\ref{sec:conclusion}. All proofs and additional theoretical results are relegated to the appendix.

%% file: ECmodel.tex
We describe the energy community structure, resources, payment and surplus functions, and the cost-causation principle in $\S$\ref{subsec:structure}-$\S$\ref{subsec:costCausation}, followed by introducing a bi-level optimization, in $\S$\ref{subsec:BilevelOpt}, that models the interaction between the community operator and its members as a Stackelberg game.

\subsection{Energy Community Structure}\label{subsec:structure}
\begin{figure}
    \centering
    \includegraphics[scale=0.5]{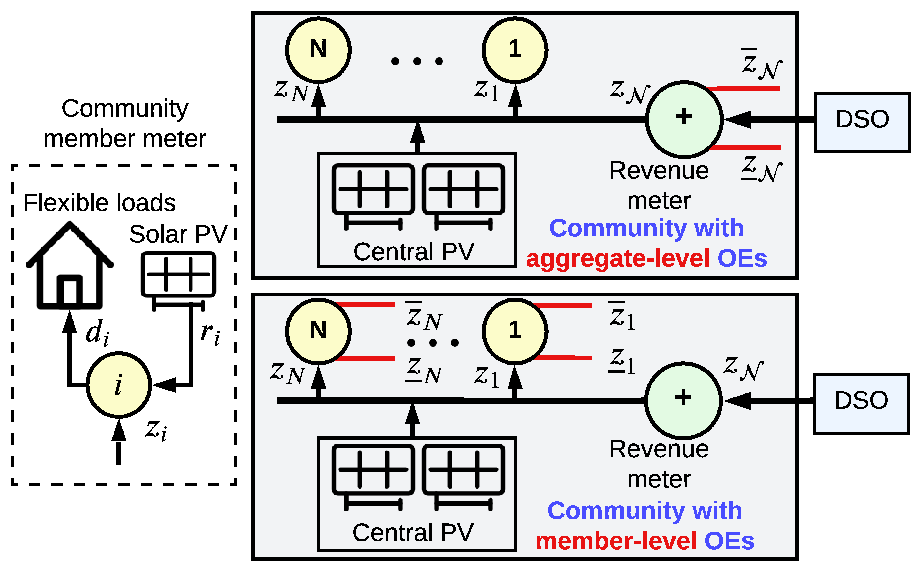}
    \vspace{-0.3cm}
    \caption{$N$-members energy community with aggregate-level OEs (top), and member-level OEs (bottom). Member consumption, renewables, and net consumption are denoted by $d_i \in \mathbb{R}_+, r_i\in \mathbb{R}_+, z_i \in \mathbb{R}$, respectively, whereas $z_{\Nc}\in \mathbb{R}$ denotes the aggregate net consumption. $\overline{z}_i\in \mathbb{R}_+, \underline{z}_i\in \mathbb{R}_-$ ($\overline{z}_{\Nc}\in \mathbb{R}_+, \underline{z}_{\Nc}\in \mathbb{R}_-$) are the member- (aggregate-) level import and export envelopes, respectively. The arrow direction indicates positive quantities.}
    \label{fig:EnergyCommunity}
\end{figure}

The profit-neutral community operator receives a single bill on behalf of its $N$ energy-sharing members, represented by the set $\Nc:=\{1,\ldots,N\}$, who are subject to the operator's market mechanism that determines the pricing and payment rules (Fig.\ref{fig:EnergyCommunity}). As depicted in Fig.\ref{fig:EnergyCommunity}, two DSO-imposed OEs models might be practiced. The first, studied in \cite{Alahmed&Cavraro&Bernstein&Tong:23AllertonArXiv},\footnote{The pricing policy under member-level OEs is in appendix \ref{sec:MemberLevelOEsPolicy}.} is when OEs are imposed at the members' revenue meters (bottom panel) to ensure safe network operation, whereas, in the second, it is adequate to impose OEs at the community revenue meter (top panel). The first model may be more prevalent in large communities with non-adjacent loads and resources. The second model is more applicable to small to medium-sized communities such as apartment buildings, schools, farms, and medical complexes. Additionally, the aggregate-level OEs model is suitable for networks where the DSO lacks full observability of customer loads and BTM DER or finds it difficult to impose OEs at the community members' meters. Under both models, the DSO-imposed OEs are communicated to the community operator {\em apriori} \cite{Liu&Ochoa&Wong&Theunissen:21TSG,Azimetal:23TSG,Alahmed&Cavraro&Bernstein&Tong:23AllertonArXiv}. Depending on the underlying network, the DSO might publish the OEs 5-minutes ahead (real-time) or 24-hours ahead (day-ahead) \cite{Liu&Ochoa&Wong&Theunissen:21TSG}.

Under both models, the operator's goal is to devise a market mechanism that maximizes the community's welfare given the DSO-imposed OEs. Community members respond to the operator's pricing by maximizing their own {\em surplus} through scheduling their local DER subject to their flexibility limits.

\subsection{Community Member DER, Payment, and Surplus}\label{subsec:resources}

\subsubsection{Community DER}\label{subsec:ModelBTM}
Every community member may have flexible loads and solar PV. The {\em consumption} vector of a bundle of $K\in \Kc:=\{1,\ldots,K\}$ devices is denoted by $\bm{d}_i \in \mathbb{R}_+^K$, which is bounded by \ie
\begin{equation}\label{eq:Conslimit}
    %0 \leq d_i \leq \overline{d}_i,~~~ \forall i\in \Nc,
  \bm{d}_i \in \mcD_i := [\underline{\bm{d}}_i, \overline{\bm{d}}_i],~~~ \forall i\in \Nc,
\end{equation}
where $ \underline{\bm{d}}_i$ and $\overline{\bm{d}}_i$ are the device bundle's lower and upper consumption limits of $i\in \Nc$, respectively. The community {\em aggregate consumption} is defined as $d_\Nc:= \sum_{i \in \Nc}  \bm{1}^\top \bm{d}_i$.

Every $i\in \Nc$ member may have a random {\em solar PV} whose output is denoted by $r_i\in \mathbb{R}_+$, which may also include the member's share from the central solar PV.\footnote{A central storage can be incorporated too, as described in \cite{Alahmed&Tong:24TEMPR}.} The community {\em aggregate generation} is defined as $r_\Nc:=\sum_{i \in \Nc} r_i$.

% The following is a deleted footnote
%\footnote{Utilizing the concept of {\em virtual NEM (VNEM)}, the community operator assigns the solar farm's output to its members through shares, and {\em virtually} accounts for the share's output as if it was a BTM DER.}

Given members' consumption and generation, the {\em net consumption} of every $i\in \Nc$ member and their {\em aggregate net consumption} are defined as 
\begin{equation}\label{eq:Netconsi}
    z_i:=  \bm{1}^\top \bm{d}_i - r_i,~~ z_\Nc:= \sum_{i\in \Nc} z_i = d_\Nc - r_\Nc,
\end{equation}
where $z_i>0$ ($z_i<0$) indicates a {\em net-consuming} ({\em net-producing}) member.

Under the {\em aggregate-level OEs} community model (Fig.\ref{fig:EnergyCommunity}), the aggregate net consumption is constrained as
\begin{equation}\label{eq:Netconslimit}
    %\underline{z}_\Nc \leq z_\Nc \leq \overline{z}_\Nc, 
z_\Nc \in \mathcal Z_\Nc := [\underline{z}_\Nc,\overline{z}_\Nc],
\end{equation}
where $\underline{z}_\Nc\leq 0$ and $\overline{z}_\Nc\geq 0$ are the export and import envelopes at the PCC, respectively. Whereas, under the {\em member-level OEs} community model (Fig.\ref{fig:EnergyCommunity}) studied in \cite{Alahmed&Cavraro&Bernstein&Tong:23AllertonArXiv}, the net consumption of every member is constrained, therefore,
\begin{equation}\label{eq:netconslimit}
z_i \in \mathcal Z_i := [\underline{z}_i,\overline{z}_i],
\end{equation}
where $\underline{z}_i\leq 0$ and $\overline{z}_i\geq 0$ are the export and import envelopes at the members' meters, respectively.\footnote{The members' OEs may be non-uniform because the DSO might discriminatively assign OEs within a customer class (e.g., residential), or because the community members do not all belong to a single customer class.} We assume the following natural relationship between {\em prosumer-level} and {\em aggregate-level} OEs
\begin{equation}
\sum_{i\in \mathcal{N}} \overline{z}_i \leq \overline{z}_\mathcal{N},~~ \sum_{i\in \mathcal{N}} \underline{z}_i \geq \underline{z}_\mathcal{N}.
\end{equation}

The OEs in (\ref{eq:Netconslimit}) and (\ref{eq:netconslimit}) are not a community operator model choice but rather imposed by the DSO.\footnote{Such constraints are the status quo, and future, in active distribution networks. See, for example, South Australia's dynamic OEs requirements \cite{AustraliaOEs:23}. }

\subsubsection{Community Members Surplus}
The profit-neutral community operator designs a pricing policy $\chi$ for its members that defines the payment function $\bm{P}^\chi(\bm{z}):= (P_1^\chi(z_1),\ldots, P_N^\chi(z_N))$, where $P_i^\chi$ is the payment function for member $i$ under $\chi$, and $\bm{z}:=(z_1,\ldots,z_N)$. The payment function may be a {\em one-part pricing}, \ie a {\em volumetric charge}, or a {\em two-part pricing}, \ie a combination of {\em volumetric} and {\em non-volumetric} (lump-sum) charges. 

The {\em surplus} of every $i \in \Nc$ community member is characterized by comfort/satisfaction from consumption and economics (payment) metrics as
\begin{equation}\label{eq:Surplusi}
    S^{\chi}_i(\bm{d}_i|r_i):=\hspace{-0.45cm} \underbrace{U_i(\bm{d}_i)}_\text{utility of consumption}\hspace{-0.2cm}-\underbrace{P^{\chi}_i(z_i)}_\text{payment under $\chi$},\hspace{-0.3cm}
\end{equation}
where $r_i$ appears through~\eqref{eq:Netconsi} and, for every $i\in \Nc$, the {\em utility of consumption function} $U_i(\bm{d}_i)$ is assumed to be additive, strongly concave, non-decreasing, and continuously differentiable with a {\em marginal utility function} $\mathbf{L}_i:=\nabla U_i=\left(L_{i 1}, \ldots, L_{i K}\right)$. We denote the {\em inverse marginal utility} vector by $\bm{f}_i:=(f_{i1},\ldots,f_{iK})$ with $f_{ik}:= L^{-1}_{ik}, \forall i\in \Nc, k\in \Kc$.

\begin{comment}
\subsubsection{Energy Community Payments and Surpluses}\label{subsec:surplusWelfare}
At the PCC, the community faces the DSO's NEM X tariff model \cite{Alahmed&Tong:22EIRACM,Alahmed&Tong:22IEEETSG}, characterized by the parameter $\pi=(\pi^+,\pi^-,\pi^0)$, which has a pricing rule $\Gamma^\pi$ and a payment rule $P^\pi$, given by
 \begin{align}
       \Gamma^\pi(z_\Nc) = \begin{cases}
\pi^+, &\hspace{-0.85em} z_\Nc\geq 0 \\ 
\pi^-, &\hspace{-0.85em} z_\Nc< 0
\end{cases},~
    P^\pi(z_\Nc) = \Gamma^\pi(z_\Nc) \cdot z_\Nc + \pi^0,\label{eq:Pcommunity}
    \end{align}
respectively, where $\pi^+\geq 0$ and $\pi^- \geq 0$ are the {\em buy} (retail) and {\em sell} (export) rates, whereas $\pi^0$ is the utility's fixed charge\footnote{For the rest of the paper, and without loss of generality, we assume the fixed charge is equal to zero $\pi^0=0$, as it does not affect the market-mechanism-induced welfare maximization.}. We assume that $\pi^+ \geq \pi^-$, which avoids risk-
free price arbitrage, given that the retail and export rates are
deterministic and known apriori.
\end{comment}

\subsection{Benchmark Prosumer: DER, Payment, and Surplus}\label{subsec:benchamrk}

The {\em benchmark} to the energy community prosumer is a standalone prosumer who faces the DSO's NEM tariff.\footnote{In the parlance of cooperative game theory, the standalone prosumer is equivalent to a singleton energy community coalition.} To ensure fairness when comparing the surplus of community members under the community pricing policy $\chi$ to their benchmark surplus under the DSO's regime, we consider that prosumer resources (BTM DER, and share from central generation) are the same with and without the community. The DSO ensures the distribution network's safe operation by imposing OEs at the benchmark customers' revenue meters, which constrain their net consumption $z_i$ 
%by ($\underline{z}_i,\overline{z}_i$) 
as in (\ref{eq:netconslimit}).

The {\em benchmark payment}, for every $i\in \Nc$, is given by the DSO's NEM X tariff, introduced in \cite{Alahmed&Tong:22IEEETSG}, as
 \begin{align}
       \pi^{\mbox{\tiny NEM}}(z_i) = \begin{cases}
\pi^+, &\hspace{-0.85em} z_i\geq 0 \\ 
\pi^-, &\hspace{-0.85em} z_i < 0
\end{cases},~
    P^{\mbox{\tiny NEM}}(z_i) = \pi^{\mbox{\tiny NEM}}(z_i) \cdot z_i,\label{eq:Pcommunity}
    \end{align}
where $\pi^+$ and $\pi^- \geq 0$ are the {\em buy} (retail) and {\em sell} (export) rates, respectively.\footnote{Here, we do not include possible fixed connection charges in the NEM X tariff, assuming that such charges are matched by membership fees.} Consistent with the practice under NEM, we assume $\pi^+ \geq \pi^-$, which also avoids risk-free price arbitrage, given that the retail and export rates are deterministic and known {\em apriori}. A NEM tariff with $\pi^-=\pi^+$ is referred to as {\em NEM 1.0}.

Similar to community members, the {\em surplus} of the {\em benchmark} prosumer, for every $i\in \Nc$, is
\begin{equation}\label{eq:SurplusBench}
S^{\mbox{\tiny NEM}}_i(\bm{d}_i|r_i):= U_i(\bm{d}_i)-P^{\mbox{\tiny NEM}}(z_i),~~ z_i:=\bm{1}^\top \bm{d}_i-r_i.
\end{equation}
The surplus-maximizing benchmark prosumer solves
\begin{align} \label{eq:BenchmarkProblem}
\Pc_{i}^{\mbox{\tiny NEM}}:   \underset{\bm{d}_i \in \mathbb{R}_+^K}{\rm maximize}~~&  S^{\mbox{\tiny NEM}}_i(\bm{d}_i|r_i):=U_i(\bm{d}_i) - P^{\mbox{\tiny NEM}}(z_i)\nn\\
\text{subject to}~~ & z_i =  \bm{1}^\top \bm{d}_{i} - r_i    \\&
%\underline{d}_i\leq d_i \leq \overline{d}_i 
\bm{d}_i \in \mcD_i, \ z_i \in \mathcal Z_i. \nonumber
%\\& \underline{z}_i \leq  z_i \leq \overline{z}_i. \nonumber
\end{align} 
 To ensure that a feasible solution to \eqref{eq:BenchmarkProblem} always exists, we assume that, for every $i \in \Nc$, the OEs ($\overline{z}_i,\underline{z}_i$) satisfy $\overline{z}_i\geq \bm{1}^\top \underline{\bm{d}}_{i} - r_i$ and $\underline{z}_i\leq \bm{1}^\top \overline{\bm{d}}_{i} - r_i$. Given this assumption, Lemma \ref{lem:BenchmarkSur} in the appendix characterizes the benchmark's maximum surplus $S^{\mbox{\tiny NEM}^\ast}_i(r_i)$ and shows that it is a function of the prosumer's renewable generation.

\subsection{Axiomatic Community Pricing}\label{subsec:costCausation}

To ensure that the community pricing rule $\chi$ is fair and just for all members, we adopt the cost-causation principle in \cite{Kirby&Milligan&Wan:06OSTI,Chakraborty&Baeyens&Khargonekar:18TPS} with some generalizations to the individual rationality axiom and generalizations due to two-part pricing. In particular, we require the pricing rule $\chi$ to satisfy Axioms \ref{ax:equity}--\ref{ax:ProfitNeutrality} below that constitute the cost-causation principle. Before describing the axioms, we introduce $\tilde{P}^\chi_i$, which denotes the volumetric part of the payment function.

\begin{axiom}[Uniform volumetric charge]\label{ax:equity}
    The pricing rule has a uniform volumetric charge if, for any two community members $i,j \in \Nc, i\neq j$, having $z_i =z_j$ results in $\tilde{P}^{\chi}_i=\tilde{P}^{\chi}_j$.
\end{axiom}
For the community pricing problem under consideration, the uniform volumetric charge corresponds to the coalition cost allocation axiom's broader {\em equal treatment of equals} premise.

\begin{axiom}[Monotonicity and cost-causation]\label{ax:monotonicity}
   For every $i\in \Nc$, the volumetric part of the payment function $\tilde{P}_i^\chi$ is monotonic and $\tilde{P}^\chi_i(0)=0$.
\end{axiom}
The monotonicity in axiom \ref{ax:monotonicity} ensures that when any two members $i,j \in \Nc, i\neq j$ have $z_i z_j \geq 0$  and $|z_i| \geq |z_j|$, then the member with higher net consumption (production) should get higher payment (compensation), \ie $|\tilde{P}^{\chi}_i(z_i)|\geq |\tilde{P}^{\chi}_j(z_j)|$. Requiring $\tilde{P}^\chi_i(0)=0$ in addition to monotonicity ensures that a member $i\in \Nc$ who causes cost, \ie $z_i>0$, is penalized for it, hence $\tilde{P}^\chi_i(z_i)\geq 0$, whereas if it mitigates cost, \ie $z_i<0$, then it is rewarded for it, hence $\tilde{P}^\chi_i(z_i)\leq 0$.

The next axiom ensures the competitiveness of the community over receiving service under the DSO's NEM X regime.

\begin{axiom}[Individual rationality]\label{ax:rationality}
   The surplus of every member $i\in \Nc$ should be no less than its benchmark, \ie $S_i^\chi \ge S_i^{\mbox{\tiny NEM}}$.
\end{axiom}

Lastly, and consistent with the {\em citizen energy community} definition under \cite{CitizenEC:EuroCommWebsite}, we require the community operator to be budget-balanced.
\begin{axiom}[Profit neutrality]\label{ax:ProfitNeutrality}
The community pricing rule must ensure the operator's profit neutrality, \ie  
\[\sum_{i\in \Nc} P^\chi_i(z_i) = P^{\mbox{\tiny NEM}}(\sum_{i\in \Nc} z_i).\]
\end{axiom}
The four axioms are used next in Definition \ref{def:CausationPrinciple}.

\begin{definition}[Cost-causation principle]\label{def:CausationPrinciple}
    A community payment function $\bm{P}^\chi$ conforms with the {\em cost-causation principle} if $\bm{P}^\chi \in \Ac:=\{\bm{P}: \text{Axioms \ref{ax:equity}--\ref{ax:ProfitNeutrality} are satisfied}\}$.
\end{definition}

Worth mentioning is that some of the celebrated {\em ex-post} allocation mechanisms such as the {\em proportional rule} and {\em Shapley value} violate the cost-causation principle \cite{Chakraborty&Baeyens&Khargonekar:18TPS}; they do not meet Axioms \ref{ax:equity}--\ref{ax:monotonicity}.

 \subsection{Bi-level Optimization}\label{subsec:BilevelOpt}
 We model the interaction between the community operator (leader) and its members (followers) as a Stackelberg game. The goal of the operator is to devise a pricing policy $\chi$ that achieves the maximum social welfare in a distributed fashion while satisfying the DSO's OEs and the generalized cost-causation principle. Every member responds to the pricing policy by maximizing their own surplus subject to their local constraints via choosing a consumption scheduling policy $\psi$.

\subsubsection{Lower Level -- Members’ Optimal Consumption Policy}
Denote member's $i$ consumption policy given the pricing policy $\chi$ by $\bm{\psi}_{i,\chi}$.
Formally, 
$$\bm{\psi}_{i,\chi}: \mathbb{R}_+^K \rightarrow \mcD_i, \ r_i \stackrel{P^\chi_i}{\mapsto} \bm{\psi}_{i,\chi}(r_i).$$
Assuming that all members are rational and surplus maximizers, the optimal consumption policy for the $i$th member $\bm{\psi}_{i,\chi}^\sharp$ (given $\chi$) is such that $\bm{\psi}^\sharp_{i,\chi}(r_i) = \bm{d}^\sharp_i$, with 
\begin{align}\label{eq:LowerLevel}
\bm{d}^\sharp_i := ~\underset{\bm{d}_i \in \mcD_i}{\operatorname{argmax}}~~&S^\chi_i(\bm{d}_i|r_i):=U_i(\bm{d}_i)-P^\chi_i(z_i)\nn\\
 		\text{subject to}~~~ &z_i= \bm{1}^\top \bm{d}_i - r_i.
\end{align}
 Note that the consumption schedule of member $i$ is based on its DER $r_i$, which comprises both private and shared resources and is stochastic.

Cooperative behaviors among the community members is beyond the scope of this work.
%\tcb{The price-taker assumption is made to rule out the formulation of price-anticipating members, who conjecture the solution   }
%\tcb{The price-taker assumption entails that members schedule their consumption as if it has no effect on the community price. If some of the members exercise price-setting behaviors, Axiom \ref{ax:rationality} of the cost-causation principle might not hold, as the additional surplus they achieve will be at the cost of other members. The price-taker assumption is more plausible when the number of community members is sufficiently large so that strategic price-setting behaviors by a member cannot improve its surplus, hence the best strategy is to be a price-taker. Limiting the amount of information revealed to members can also limit price-setting behaviors \cite{Samadi&MohsenianEtal:12TSG}.}

 \subsubsection{Upper Level -- Operator's Optimal Pricing Policy}
 The operator's pricing objective is to maximize the expected community welfare defined by the total surplus of its members

\begin{equation*}
W^{\chi,\bm{\psi}_{\chi}}:=\sum_{i\in \Nc} \mbbE[S_i^{\chi}(\bm{\psi}_{i,\chi}(r_i),r_i)],
\end{equation*}
where the expectation is taken over member DERs ($r_i$), and $\bm{\psi}_{\chi}:=\{\bm{\psi}_{1,\chi},\ldots,\bm{\psi}_{N,\chi}\}$. Under Axiom \ref{ax:ProfitNeutrality}, the (conditional) welfare maximizing pricing policy $\chi^{\sharp}_{\bm{\psi}}$  (given $\bm{\psi}$) is defined by $\chi^{\sharp}_{\bm{\psi}}: \mathbb R_+^N \rightarrow \mathbb R_+, \ \bm{r} \mapsto P^{\chi},$ where

\begin{align}\label{eq:UpperLevel}
 P^{\chi} := \underset{P(\cdot) \in \Ac}{\operatorname{argmax}}& \Bigg(W^{\chi_{\bm{\psi}}} = \mbbE\Big[\sum_{i\in \Nc} U_i(\bm{\psi}^\sharp_{i,\chi}(r_i)) - P^{\mbox{\tiny NEM}}(z_{\Nc})\Big]\Bigg)\nn\\
 \text{subject to}&~~~ z_{\Nc} = \sum_{i\in \Nc} \big(\bm{1}^\top \bm{\psi}^\sharp_{i,\chi}(r_i)- r_i\big)\\
 &~~~	z_{\Nc} \in \mathcal{Z}_\Nc.\nn
\end{align}
The infinite-dimensional program in (\ref{eq:UpperLevel}) cannot be solved easily without exploiting the NEM tariff structure.
\subsubsection{Stackelberg Equilibrium}
The bi-level optimization above defines the so-called {\em Stackelberg equilibrium} (or Stackelberg strategy).  %Specifically, ($\chi^\ast,\bm{\psi}^\ast$) is a Stackelberg equilibrium if (a) for all $\chi \in \Xc$ and $i\in \Nc$,  $ S^\chi_i(\psi^\ast_i (r_i),r_i) \ge S^\chi_i(\psi_i (r_i),r_i)$ for all $\bm{\psi} \in \Psi$; (b) for all $\bm{\psi} \in \Psi, W^{\chi^\ast,\bm{\psi}^\ast} \ge \mbbE[\sum_i S_i^\chi(\psi^\ast_i(r_i),r_i)]$.  
Specifically, ($\chi^\ast,\bm{\psi}^\ast$) is a Stackelberg equilibrium since (a) for all $\chi \in \Xc$ and $i\in \Nc$,  $ S^\chi_i(\bm{\psi}^\ast_i (r_i),r_i) \ge S^\chi_i(\bm{\psi}_i (r_i),r_i)$ for all $\bm{\psi} \in \Psi$; (b) for all $\bm{\psi} \in \Psi, W^{\chi^\ast,\bm{\psi}^\ast} \ge \mbbE[\sum_i S_i^\chi(\bm{\psi}^\ast_i(r_i),r_i)]$. Specifically, the Stackelberg equilibrium is the optimal community pricing when community members optimally respond to the community pricing.

%% file: MktMech.tex
In this section, we present the community pricing policy and its structural properties, followed by showing the optimal response of members, and the corresponding community market equilibrium to the bi-level optimization in (\ref{eq:LowerLevel})-(\ref{eq:UpperLevel}).

\subsection{OEs-Aware Pricing Policy}\label{subsec:OEawarePricing}
The operator announces the community pricing policy given the aggregate renewable DG $r_{\Nc}$ as shown next. For every $i\in \Nc$, the OEs-aware energy community pricing policy consists of two parts, as 
\begin{equation}\label{eq:TwoPartPricing}
\chi^{\ast}: \bm{r} \mapsto P^{\chi^\ast}_i(z_i)=\underbrace{\pi^\ast(r_\Nc)\cdot z_i}_\text{volumetric charge} - \underbrace{A_i^\ast(r_\Nc)}_\text{non-volumetric charge},
\end{equation}
where the price $\pi^\ast(r_\Nc)$ and the non-volumetric charge $A_i^\ast(r_\Nc)$ are defined next.

\begin{policy1*}
For every $i\in \Nc$, the price obeys a four thresholds policy on $r_\Nc$, as
\begin{align}
\pi^{\ast}(r_{\Nc})=\begin{cases}   \chi^+(r_{\Nc})&\hspace{-.10em}, r_{\Nc} \leq \sigma_1 \\
\pi^+&\hspace{-.10em}, r_{\Nc} \in (\sigma_1, \sigma_2) \\ \chi^z(r_{\Nc}) &\hspace{-.10em}, r_{\Nc} \in[\sigma_2,\sigma_3] \\ \pi^-& \hspace{-.10em},
 r_{\Nc} \in(\sigma_3,\sigma_4)\\
 \chi^-(r_{\Nc})& \hspace{-.10em}, r_{\Nc} \geq \sigma_4,
 \end{cases}\label{eq:PricingMechanism}
 \end{align}
 where the four thresholds $\sigma_1 \leq \sigma_2 \leq \sigma_3 \leq \sigma_4$ are defined as
 %
% \begin{equation}
% \begin{split}
% \sigma_1 &:= \sigma_2 - \overline{z}_\Nc, \\
% \sigma_3 &:=\sum_{i \in \Nc} \bm{1}^\top [\bm{f}_{i}%(\bm{1}\pi^-)]^{\overline{\bm{d}}_{i}}_{\underline{\bm{d}}_{i}}\nn
 %\end{split}
% \quad \quad
% \begin{split}
%     \sigma_2 &:=\sum_{i \in \Nc} \bm{1}^\top [\bm{f}_{i}(\bm{1}\pi^+)]^{\overline{\bm{d}}_{i}}_{\underline{\bm{d}}_{i}}\nn\\
%\sigma_4 &:= \sigma_3 - \underline{z}_\Nc,\nn
% \end{split}
%\end{equation}
 %
\begin{equation}
\begin{split}
\sigma_1 &:= \sigma_2 - \overline{z}_\Nc, \\
\sigma_3 &:=\sum_{i \in \Nc} \bm{1}^\top [\bm{f}_{i}(\bm{1}\pi^-)]^{\overline{\bm{d}}_{i}}_{\underline{\bm{d}}_i}\nn
\end{split}
\quad \quad
\begin{split}
\sigma_2 &:=\sum_{i \in \Nc} \bm{1}^\top [\bm{f}_{i}(\bm{1}\pi^+)]^{\overline{\bm{d}}_{i}}_{\underline{\bm{d}}_i}\nn\\
\sigma_4 &:= \sigma_3 - \underline{z}_\Nc,\nn
\end{split}
\end{equation}
and the prices $\chi^+(r_{\Nc}):=\mu^\ast_1(r_{\Nc})$, $\chi^z(r_{\Nc}):=\mu^\ast_2(r_{\Nc})$, $\chi^-(r_{\Nc}):=\mu^\ast_3(r_{\Nc})$ are, respectively, the solutions of:
% \begin{align}\label{eq:MonotonicPrices1}
% \sum_{i \in \Nc} \bm{1}^\top [\bm{f}_{i}(\bm{1}\mu_1)]^{\overline{\bm{d}}_{i}}_{\underline{\bm{d}_{i}}} &= r_{\Nc}+ \overline{z}_\Nc\\\label{eq:MonotonicPrices2}
% \sum_{i \in \Nc} \bm{1}^\top [\bm{f}_{i}(\bm{1}\mu_2)]^{\overline{\bm{d}}_{i}}_{\underline{\bm{d}_{i}}} &= r_{\Nc}\\\label{eq:MonotonicPrices3}
% \sum_{i \in \Nc} \bm{1}^\top [\bm{f}_{i}(\bm{1}\mu_3)]^{\overline{\bm{d}}_{i}}_{\underline{\bm{d}_{i}}} &= r_{\Nc} + \underline{z}_\Nc,
%\end{align}
%
\begin{align}\label{eq:MonotonicPrices1}
\sum_{i \in \Nc} \bm{1}^\top [\bm{f}_{i}(\bm{1}\mu_1)]^{\overline{\bm{d}}_{i}}_{\underline{\bm{d}}_i} &= r_{\Nc}+ \overline{z}_\Nc\\\label{eq:MonotonicPrices2}
\sum_{i \in \Nc} \bm{1}^\top  [\bm{f}_{i}(\bm{1}\mu_2)]^{\overline{\bm{d}}_{i}}_{\underline{\bm{d}}_i} &= r_{\Nc}\\\label{eq:MonotonicPrices3}
\sum_{i \in \Nc} \bm{1}^\top [\bm{f}_{i}(\bm{1}\mu_3)]^{\overline{\bm{d}}_{i}}_{\underline{\bm{d}}_i} &= r_{\Nc} + \underline{z}_\Nc.
\end{align}
\end{policy1*}

Fundamentally, the price $\pi^\ast(r_{\Nc})$ is used to induce a welfare-maximizing members' response. When $r_\Nc \in (\sigma_1,\sigma_2)$ and $r_\Nc \in (\sigma_3,\sigma_4)$, the operator directly passes the DSO's NEM prices $\pi^+$ and $\pi^-$, respectively. The dynamic prices $\chi^+(r_{\Nc})$ and $\chi^-(r_{\Nc})$ are the Lagrange multipliers associated with the import and export OEs constraints, respectively, whereas the dynamic price $\chi^z(r_{\Nc})$ is the Lagrange multiplier of the community's energy-balancedness.

The following proposition describes the order of the five prices alongside their monotonicity.
\begin{proposition}\label{corol:PriceOrder}
    The OEs-aware prices are ordered as
$$\chi^+(r_\Nc)\geq \pi^+ \geq \chi^z (r_\Nc) \geq \pi^- \geq \chi^-(r_\Nc)\geq 0,$$
and the prices $\chi^+(r_\Nc), \chi^z(r_\Nc)$ and $\chi^-(r_\Nc)$ are monotonically decreasing with $r_\Nc$.
\end{proposition}
\noindent {\em Proof:} See Appendix \ref{sec:AppProofs}.\hfill$\Box$\\
We discuss, in $\S$\ref{subsec:StructureIntuition}, the structure and intuitions of the OEs-aware pricing policy in greater detail.

\begin{figure}
    \centering
    \includegraphics[scale=0.49]{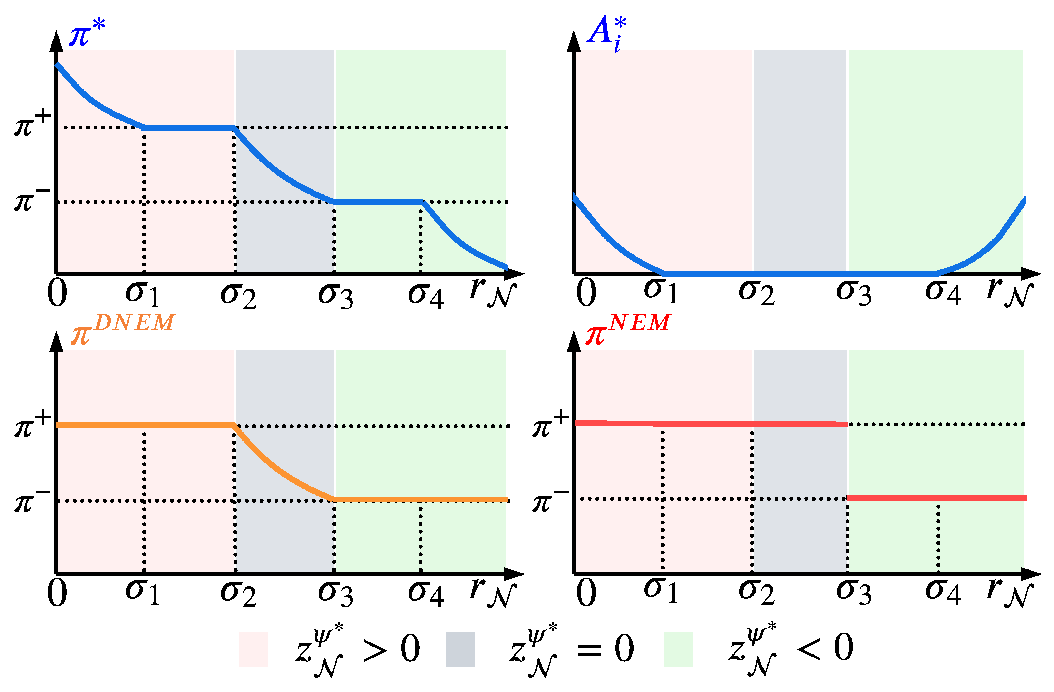}
    \vspace{-0.47cm}
    \caption{Community pricing policy with OEs (blue) and without OEs \cite{Alahmed&Tong:24TEMPR} (yellow), compared to the DSO NEM X (red) under optimal community members' response shown in Lemma \ref{lem:OptSchedule}, all with respect to aggregate renewables $r_{\Nc}$. }
    \label{fig:MktMech}
\end{figure}

A two-part pricing is necessary because, without the non-volumetric charge, the operator is not {\em profit-neutral}; a volation to Axiom \ref{ax:ProfitNeutrality}. In fact, without a non-volumetric charge, the operator is {\em revenue-adequate}. The revenue-adequacy arises from the price difference between $\pi^{\mbox{\tiny NEM}}(z_{\Nc})$ and $\pi^\ast(r_{\Nc})$ when the OEs bind. Indeed, mathematically, the payment difference is given by
\begin{align}
   \sum_{i\in \Nc}\hspace{-0.1cm} \tilde{P}^{\chi^\ast}_i\hspace{-0.1cm}(z_i) - P^{\mbox{\tiny NEM}}(\sum_{i\in \Nc} z_i)\hspace{-0.1cm} &= \begin{cases}
(\chi^+(r_{\Nc})-\pi^+) \overline{z}_\Nc &\hspace{-0.25cm} ,r_{\Nc}\leq \sigma_1 \\ 
(\chi^-(r_{\Nc})-\pi^-) \underline{z}_\Nc &\hspace{-0.25cm}, r_{\Nc}\geq \sigma_4\\
0&\hspace{-0.25cm}, \text{o.w}
\end{cases}\label{eq:PaymentDifference}\\
& \stackrel{\text{(A)}}{\geq} 0, \nonumber
\end{align}
where (A) is straightforward from Proposition \ref{corol:PriceOrder}. In words, (\ref{eq:PaymentDifference}) implies that the operator collects money from its members higher than what it pays to the DSO.
To neutralize the operator's profit, we propose a non-volumetric charge that re-distributes the profit based on the {\em proportional allocation.}\footnote{Such profit distribution is similar to the revenue adequacy result of LMP in wholesale electricity markets, which is neutralized by {\em merchandising surplus}.}
\begin{policy2*}
    For every $i\in \Nc$, the non-volumetric charge obeys a two thresholds policy on $r_\Nc$, as
    \begin{equation}\label{eq:fixedreward}
        A^\ast_i(r_{\Nc})= \begin{cases}
(\chi^+(r_{\Nc})-\pi^+) \left(\overline{z}_i + \frac{\overline{z}_\Nc-\sum_{i\in \Nc}\overline{z}_i}{N}\right) &\hspace{-0.33cm} ,r_{\Nc}\leq \sigma_1 \\ 
(\chi^-(r_{\Nc})-\pi^-) \left(\underline{z}_i + \frac{\underline{z}_\Nc-\sum_{i\in \Nc}\underline{z}_i}{N}\right) &\hspace{-0.33cm}, r_{\Nc}\geq \sigma_4\\
0&\hspace{-0.43cm}, \text{o.w}.
\end{cases} 
    \end{equation}
\end{policy2*}

As will be shown in Lemma \ref{lem:CostCausation}, the proposed non-volumetric charge is crucial not only in achieving {\em profit-neutrality} (Axiom \ref{ax:ProfitNeutrality}), but also {\em individual rationality} (Axiom \ref{ax:rationality}). Given that there might be other allocation rules to distribute the profit, while not violating the pricing axioms, one could use the price in (\ref{eq:PricingMechanism}) and construct a profit-sharing coalitional game to distribute the operator profit to the coalition (community) members.

\subsection{OE-Aware Pricing Policy: Structure and Intuitions}\label{subsec:StructureIntuition}
We discuss next the structural properties and intuitions of the OEs-aware pricing policy, depicted in Fig.\ref{fig:MktMech}, and show how it differs from the pricing policy when there are no OEs \cite{Alahmed&Tong:24TEMPR} and when the OEs are at the member-level \cite{Alahmed&Cavraro&Bernstein&Tong:23AllertonArXiv}.

%\subsubsection{Two-Part Pricing Policy Structure}

%\begin{enumerate}[leftmargin=*,label=(\alph*)]

\emph{a)} Two-part pricing: The pricing rule $P^{\chi^\ast}_i(\cdot)$ charges or compensates members based on a two-part pricing. The volumetric charge is the product of the dynamic price in (\ref{eq:PricingMechanism}) and the member's net consumption $z_i$. The non-volumetric part (\ref{eq:fixedreward}) is a reward ($A_i^\ast \geq 0$) paid to each member whenever $r_{\Nc} \notin (\sigma_1,\sigma_4)$.
The reward each member gets is proportional to their benchmark's OEs $\underline{z}_i,\overline{z}_i$ plus any benefit due to aggregation ($\overline{z}_\Nc- \sum_{i\in \Nc} \overline{z}_i, \underline{z}_\Nc- \sum_{i\in \Nc} \underline{z}_i$). 

\emph{b)} Four-thresholds and resource-aware policy: The structure of the dynamic price $\pi^\ast$ obeys four renewable-independent thresholds that depend only on the DSO's NEM X rates, OEs, and members' willingness to consume.  The price and non-volumetric rewards are both announced based on $r_{\Nc}$ (Fig.\ref{fig:MktMech}), with the price $\pi^\ast(r_{\Nc})$ being a monotonically decreasing function of $r_{\Nc}$. The payment transitions from two-part to one-part and then to two-part again, as $r_{\Nc}$ increases. The payment function is two-part only when one of the OEs binds.

\emph{c)} Minimal member information: Within our framework, private member information specifically refers to its utility, marginal utility, and demand functions. The mechanism preserves members' privacy by not requiring the knowledge of the functional forms of their private information. As shown in (\ref{eq:PricingMechanism}) and (\ref{eq:fixedreward}), to announce the community dynamic price and fixed reward, the operator needs to compute i) the four thresholds $\sigma_1-\sigma_4$, which are then compared to the aggregate renewables in the community, and ii) the prices $\chi^+(r_{\Nc}), \pi^+, \chi^z(r_{\Nc}), \pi^-, \chi^-(r_{\Nc})$. The computation of both (i) and (ii) requires the aggregate and prosumer-level OEs ($\overline{z}_\Nc,\underline{z}_\Nc, \overline{z}_i,\underline{z}_i$), which are already known to the operator.

The four thresholds can be computed without compromising members' privacy. In particular, because NEM X rates ($\pi^+,\pi^-$) are public, each member $i$ provides a value for $[\bm{f}_{i}(\bm{1}\pi^+)]^{\overline{{\bm{d}}}_{i}}_{\underline{{\bm{d}}}_{i}}$ and $[\bm{f}_{i}(\bm{1}\pi^-)]^{\overline{{\bm{d}}}_{i}}_{\underline{{\bm{d}}}_{i}}$ {\em apriori}, from which the operator computes $\sigma_1-\sigma_4$.

Computing the dynamic prices $\chi^+(r_{\Nc}), \chi^z(r_{\Nc}), \chi^-(r_{\Nc})$, which are bounded by the NEM X rates ($\pi^+,\pi^-$) follows the same method, but requires the provision of a range of values from members rather than a single value. The operator asks the members to provide a value for each price sample within the price range, which are then used to compute (\ref{eq:MonotonicPrices1})-(\ref{eq:MonotonicPrices3}) using $r_\Nc$ and the aggregate-level OEs. The key to preserving the actual demand functions and the identity of the members, in this case, is that to compute these dynamic prices, the operator only needs the {\em aggregate} of members' values rather than individual data points. To this end, a direct application of a standard {\em differential privacy} algorithm \cite{Dwork&Roth:14MonographNOW} achieves a provable level of ($\epsilon-\delta$) differential privacy. Therefore, what can be learned about any member $i$ from a differentially private computation is effectively restricted to what can be learned about the member $i$ from all other $\Nc \backslash i$ members' data if $i$'s own data is not included in the computation \cite{Dwork&Roth:14MonographNOW}.

\emph{d)} Non-discriminatory volumetric charge: The price $\pi^\ast$ is uniform to all members, and it is always within NEM X price range except when $r_{\Nc}\notin (\sigma_1,\sigma_4)$. Also, unlike the DSO's NEM tariff in (\ref{eq:Pcommunity}), the operator charges its members a price that is not contingent on the member's net consumption $z_i$.

If the DSO's OEs are non-discriminatory, \ie $\underline{z}_i = \underline{z}, \overline{z}_i = \overline{z}, \forall i \in \Nc$, the community's non-volumetric charge also becomes non-discriminatory (uniform), as 
    $$A^\ast_i(r_{\Nc})= \begin{cases}
(\chi^+(r_{\Nc})-\pi^+) \overline{z}_\Nc/N & ,r_{\Nc}\leq \sigma_1 \\ 
(\chi^-(r_{\Nc})-\pi^-) \underline{z}_\Nc/N &, r_{\Nc}\geq \sigma_4\\
0&, \text{o.w}.
\end{cases} $$
The distribution of the accrued profit above is simply based on the {\em equal division allocation}.

\emph{e)} Endogenously-determined market roles: In conventional electricity markets, without storage, the roles of agents as {\em buyers} or {\em sellers} are predetermined. For example, in the day-ahead electricity market, participants must declare their roles through bids and offers, which are then aggregated by the operator to come up with a price that clears the market based on the objective. %making the market roles exogenous to the clearing price. 
In contrast to conventional electricity markets and to the market mechanisms in \cite{Vespermann&Hamacher&Kazempour:21TPS,Kalathil&Wu&Poolla&Varaiya:19TSG,Morstyn&Teytelboym&Mcculloch:19TSG} that considers cost re-allocation, under the proposed pricing policy, the roles of community members are endogenously determined by the community price $\pi^\ast$. That is, given the price, community members are liberated to choose their roles. Specifically, after the community price is announced, the member's local renewables and demand bid will determine whether it will assume a buyer or seller role. This is further illustrated in $\S$\ref{subsec:MemberProblem}, which shows how community members determine their consumption schedule given $\chi$.

% Once you have storage, the argument is not true.

As shown in Fig.\ref{fig:MktMech}, the price lies between the DSO's NEM X retail $\pi^+$ and export $\pi^-$ rates when $r_{\Nc} \in (\sigma_1, \sigma_4)$. In particular, the community directly passes the DSO prices $\pi^+$ and $\pi^-$ when $r_{\Nc}\in (\sigma_1,\sigma_2)$ and $r_{\Nc}\in (\sigma_3,\sigma_4)$, respectively, and charge a price that dynamically varies between $\pi^+$ and $\pi^-$ depending on $r_{\Nc}$ when $r_{\Nc}\in [\sigma_2,\sigma_3]$ (similar to D-NEM without OEs \cite{Alahmed&Tong:24TEMPR}, yellow curve in Fig.\ref{fig:MktMech}, and with OEs \cite{Alahmed&Cavraro&Bernstein&Tong:23AllertonArXiv}, Appendix \ref{sec:MemberLevelOEsPolicy}). When $r_{\Nc}$ is relatively low ($r_{\Nc} \leq \sigma_1$) and the import OE is binding, the operator charges a price higher than $\pi^+$ to induce its members to reduce consumption, ensuring that $z_\Nc=\overline{z}_\Nc$. On the other hand, when $r_{\Nc}$ is relatively high ($r_{\Nc} \geq \sigma_4$) and the export OE is binding, the operator charges a price below $\pi^-$ to behoove its members to increase consumption, ensuring that $z_\Nc=\underline{z}_\Nc$.

The intuition of the non-volumetric reward $A^\ast_i$ is that because the operator charges a higher price $\chi^+(r_{\Nc})\geq \pi^+$ when $r_{\Nc}$ is low $r_{\Nc}\leq \sigma_1$, and charges a lower price $\chi^-(r_{\Nc})\leq \pi^-$ when $r_{\Nc}$ is high $r_{\Nc}\geq \sigma_4$, it compensates its members for the losses in the volumetric charge through the non-volumetric rewards. The reward each customer gets $A^\ast_i$ is proportional to their benchmark's OEs $\underline{z}_i,\overline{z}_i$ because the operator needs to better reward customers with less restrictive OEs to ensure that they have enough incentive to join the community, which is required to satisfy the individual rationality axiom.

%In the next section, we formulate the community member problem, and corresponding optimal decisions and surplus levels, which enable assessing individual rationality under the proposed pricing policy.

\subsubsection{Comparison to Member-Level OEs}
We highlight two key differences between the pricing mechanism of a community with aggregate-level and member-level OEs (Appendix \ref{sec:MemberLevelOEsPolicy}). First, under the aggregate-level OEs framework, the pricing policy induces the members toward a collective response that abides by the OEs, whereas under member-level OEs, the price ($\pi^{\mbox{\tiny DNEM}}(\bm{r})$ in \ref{eq:MemberLevelPricingMechanism}) is announced and members must account for their local OEs when scheduling their consumption. This fundamental difference resulted in a rather simpler policy in the member-level OEs framework, which abides by one-part pricing with a two-threshold ($\Theta_1(\bm{r}),\Theta_2(\bm{r})$) price that is bounded between the NEM X prices, \ie $\pi^{\mbox{\tiny DNEM}}(\bm{r}) \in [\pi^-,\pi^+]$. Second, to compute the price, the member-level OEs community operator needs full information about community renewable DG $\bm{r}$, whereas in the aggregate-level OEs framework, only the aggregate DG $r_{\Nc}$ is needed. 
%Theorem \ref{thm:AgglevelVSMemLevel} establishes the somewhat counterintuitive result that the surplus of a member under aggregate-level OEs is always no less than the surplus of a member under member-level OEs.

\subsection{Community Member Problem and Optimal Decisions}\label{subsec:MemberProblem}
Given $r_{\Nc}$, the community price is announced, and accordingly every member $i\in \Nc$ solves for the optimal consumption policy 
$\bm{\psi}^\ast_{i,\chi^\ast}: \mathbb{R}_+^K \rightarrow \mcD_i, \ r_i \stackrel{{P^{\chi^\ast}_i}}{\mapsto} \bm{d}_i^{\psi^\ast}:= \bm{\psi}^\ast_i(r_i),$
where
\begin{align} \label{eq:argmax_d+}
\bm{d}_i^{\psi^\ast}=& \underset{\bm{d}_i \in \mcD_i}{\operatorname{argmax}}~~  S_i^{\chi^\ast}(\bm{d}_i|r_i):=U_i\left(\bm{d}_i\right)- P^{\chi^\ast}_i(z_i) \nonumber \\
& \text { subject to } \quad~ z_i = \bm{1}^\top \bm{d}_i-r_i.
\end{align} 
The following lemma characterizes the optimal consumption and net consumption of the members, and the community.

\begin{lemma}[Member optimal consumption]\label{lem:OptSchedule}
    For every member $i\in \Nc$, given the pricing policy, the maximum surplus $S^{\chi^\ast}_i(\bm{\psi}^\ast_i(r_i),r_i)$
    is achieved by the optimal consumption and net consumption given, respectively, by
    \begin{align}
        \bm{d}^{\psi^\ast}_{i}(\pi^\ast) = [\bm{f}_{i}(\bm{1}\pi^\ast)]_{\underline{\bm{d}}_{i}}^{\overline{\bm{d}}_{i}},~~
        z^{\psi^\ast}_i(\pi^\ast) = \bm{1}^\top \bm{d}^{\psi^\ast}_{i}(\pi^\ast) - r_i. \label{eq:MemberOptz}
    \end{align}
\end{lemma}
\noindent {\em Proof:} See Appendix \ref{sec:AppProofs}.\hfill$\Box$\\
%and given the monotonicity of $f_{ik}(\cdot)$ and $\pi^\ast$, shows that the member's optimal consumption $d^{\psi^\ast}_{ik}$ is a monotonically increasing function of $r_{\Nc}$, leading to a monotonically decreasing optimal net consumption $z^{\psi^\ast}_i$ with $r_{\Nc}$. 
Given $r_\Nc$, and the optimal schedule of every member in
Lemma \ref{lem:OptSchedule}, the community aggregate consumption is  $d^{\psi^\ast}_\Nc(\pi^\ast)\hspace{-0.1cm} =\hspace{-0.1cm} \sum_{i\in \Nc}  \bm{1}^\top \bm{d}^{\psi^\ast}_{i}(\pi^\ast)$ and aggregate net consumption is
\begin{align}
       z^{\psi^\ast}_{\Nc}(\pi^\ast) =\hspace{-0.13cm}\sum_{i\in \Nc} z^{\psi^\ast}_i(\pi^\ast) \hspace{-0.10cm}= \begin{cases}
           \overline{z}_\Nc &\hspace{-0.2cm},  r_{\Nc} \leq \sigma_1\\
          d^{\psi^\ast}_\Nc(\pi^+) - r_{\Nc}&\hspace{-0.2cm},  r_{\Nc} \in (\sigma_1,\sigma_2)\\
           0&\hspace{-0.2cm},  r_{\Nc} \in [\sigma_2,\sigma_3]\\
            d^{\psi^\ast}_\Nc(\pi^-) - r_{\Nc}&\hspace{-0.2cm},  r_{\Nc} \in (\sigma_3,\sigma_4)\\
            \underline{z}_\Nc&\hspace{-0.2cm},  r_{\Nc} \geq \sigma_4.
       \end{cases}\label{eq:AggOptz}
   \end{align}

\begin{figure}
    \centering
    \includegraphics[scale = 0.44]{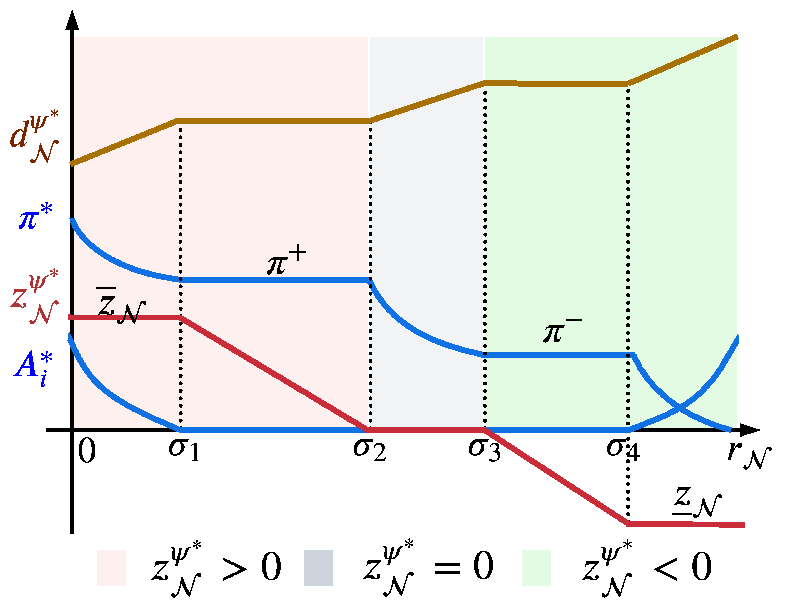}
    \vspace{-0.34cm}
    \caption{Optimal aggregate consumption (brown), net consumption (red), and the OEs-aware pricing policy (blue) with respect to aggregate renewables $r_{\Nc}$.}
    \label{fig:OptConsNetCons}
\end{figure}

Fig.\ref{fig:OptConsNetCons} depicts the optimal aggregate consumption $d^{\psi^\ast}_\Nc$ and net consumption $z^{\psi^\ast}_\Nc$, which define the community operational zones. One can see the monotonicity of $z^{\psi^\ast}_\Nc$ in $r_{\Nc}$ and that the prices $\chi^+(r_{\Nc}), \chi^-(r_{\Nc})$ are essentially dual variables of the import, and export OE constraints, respectively. $\chi^z(r_{\Nc})$ is the dual variable of energy balancing the community. The community is net-consuming when $r_{\Nc}\leq \sigma_2$, net-producing when $r_{\Nc} \geq \sigma_4$, and energy-balanced when $r_{\Nc} \in [\sigma_2,\sigma_3]$.

%Note how the dynamically decreasing price when $r_{\Nc} \geq \sigma_4$ incentivized the members to increase their consumption, which enabled the community to meet the export OE even at higher $r_{\Nc}$. On the other hand, the dynamically increasing price with decreasing $r_{\Nc} \leq \sigma_1$, induced the members to reduce their consumption, which enabled the community to meet the import OE.

\subsection{Stackelberg Equilibrium and Properties}\label{subsec:StackEq}
Given the optimal response of the benchmark (Lemma 2 in \cite{Alahmed&Cavraro&Bernstein&Tong:23AllertonArXiv}) and the community member (Lemma \ref{lem:OptSchedule}), and the pricing policy in $\S$\ref{subsec:OEawarePricing}, the following Lemma \ref{lem:CostCausation} establishes {\em cost causation conformity} of the OEs-aware pricing policy.

\begin{lemma}[Cost-causation conformity]\label{lem:CostCausation}
The OEs-aware pricing policy $\chi^\ast$ conforms with the cost-causation principle, \ie $\bm{P}^{\chi^\ast}\in \Ac$.
 \end{lemma}
 \noindent {\em Proof:} See Appendix \ref{sec:AppProofs}.\hfill$\Box$\\
 A major result from Lemma \ref{lem:CostCausation} is that every member under the proposed pricing policy
 achieves a surplus no less than its benchmark, \ie $
  S^{\chi^\ast}_i(\bm{\psi}^\ast_i(r_i),r_i)  \geq S^{\mbox{\tiny NEM}^\ast}_i(r_i),$
where $S^{\mbox{\tiny NEM}^\ast}_i(r_i)$ is the optimal surplus of the benchmark (Lemma 2 in \cite{Alahmed&Cavraro&Bernstein&Tong:23AllertonArXiv}).

Theorem \ref{thm:Equilibrium}  below establishes the Stackelberg equilibrium under the proposed OEs-aware pricing policy.

\begin{theorem}[Equilibrium and welfare optimality]\label{thm:Equilibrium}
The solution ($\chi^\ast,\psi^\ast$) is an equilibrium to (\ref{eq:LowerLevel})-(\ref{eq:UpperLevel}). Also, the equilibrium solution achieves the community maximum social welfare under centralized operation, \ie
    \begin{align} \label{eq:SurplusMax_model2}
(\bm{d}_1^{\psi^\ast},\ldots,\bm{d}_N^{\psi^\ast}) = \underset{(\bm{d}_1,\ldots,\bm{d}_N)}{\rm argmax}&~~ \mathbb{E}\big[\sum_{i\in \Nc} U_i(\bm{d}_i)-P^{\mbox{\tiny NEM}}(z_\Nc)\big]\nn \\ \text{subject to} &~~ z_\Nc = \sum_{i\in \Nc} \big(\bm{1}^\top \bm{d}_{i}-r_i\big) \nn\\&\hspace{0.4cm} z_{\Nc} \in \mathcal Z_\Nc, \bm{d}_i \in \mcD_i ~ \forall i\in \Nc.\nn
\end{align}
\end{theorem}
\noindent {\em Proof:} See Appendix \ref{sec:AppProofs}.\hfill$\Box$\\
Theorem \ref{thm:Equilibrium}'s proof leverages Lemmas \ref{lem:OptSchedule}--\ref{lem:CostCausation} above, Lemma 2 in \cite{Alahmed&Cavraro&Bernstein&Tong:23AllertonArXiv} that characterizes the maximum surplus of the benchmark and Lemma \ref{lem:CentralizedWelfare} that solves for the community maximum welfare under centralized operation.

%% file: VoC.tex
We showed in Lemma \ref{lem:CostCausation} that for every member under the OEs-aware pricing policy, joining the community is advantageous, as it achieves surplus levels no less than the maximum surplus of the benchmark. To quantify and analyze this added benefit after joining the community, we introduce the {\em value of community (VoC)} metric, defined as the aggregate expected difference between community members' surplus under the OEs-aware pricing policy and the benchmark surplus\footnote{VoC definition is analogous to the {\em value of storage} definition in \cite{Xu&Tong:17TAC}.}
\begin{equation}
\mbox{VoC}^{\chi^\ast,\mbox{\tiny NEM}}\hspace{-0.1cm}:= \hspace{-0.15cm} \sum_{i \in \Nc}\hspace{-0.1cm} \mbox{VoC}_i^{\chi^\ast,\mbox{\tiny NEM}}\hspace{-0.15cm}:=\hspace{-0.1cm} \sum_{i \in \Nc} \hspace{-0.1cm}
\mbbE  \hspace{-0.08cm}\left[S^{\chi^\ast}_i\hspace{-0.1cm}(\bm{\psi}^\ast_i(r_i),r_i)\hspace{-0.1cm} - S^{\mbox{\tiny NEM}^\ast}_i\hspace{-0.1cm}(r_i)\right],   \label{eq:VoCdef}
\end{equation}
where $\mbox{VoC}_i^{\chi^\ast,\mbox{\tiny NEM}}$ is the value gained by member $i$ after joining the community,\footnote{There are still un-captured values for joining an energy community, such as wider DER accessibility and improved economies of scale.} and the expectation is taken over $r_{i}$. From Lemma \ref{lem:CostCausation}, we know that $\mbox{VoC}_i^{\chi^\ast,\mbox{\tiny NEM}}\geq 0$. Corollary \ref{corol:PositiveVoC} below compares the benchmark prosumer's surplus given in Lemma 2 in \cite{Alahmed&Cavraro&Bernstein&Tong:23AllertonArXiv} to the member surplus under equilibrium.

\begin{figure}
    \centering
    \includegraphics[scale = 0.42]{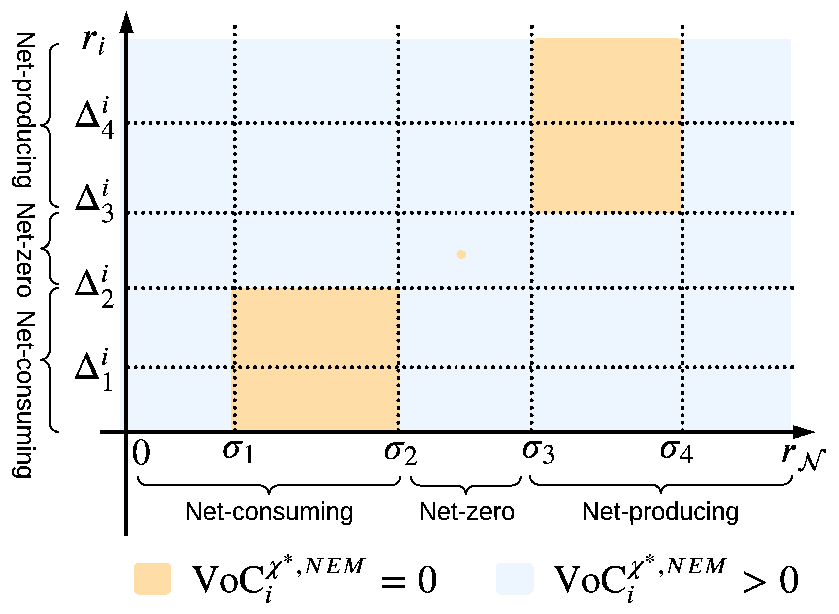}
    \vspace{-0.35cm}
    \caption{VoC of every member (Corollary \ref{corol:PositiveVoC}). $\Delta_1^i-\Delta_4^i$ are the thresholds of prosumer's $i$ benchmark optimal consumption policy.}
    \label{fig:CorolDepiction}
\end{figure}

\begin{corollary}[Member and community net-consumption complementarity]\label{corol:PositiveVoC}
    For every member $i \in \Nc$ under the OEs-aware pricing policy, if $\pi^-<\pi^+$, $\sum_{i\in \Nc} \overline{z}_i < \overline{z}_\Nc$ and $\sum_{i\in \Nc} \underline{z}_i > \underline{z}_\Nc$, then $\mbox{VoC}^{\chi^\ast,\mbox{\tiny NEM}}_i>0$ except in the following three cases
    \begin{enumerate}[label=(\roman*)]
        \item $r_\Nc\in [\sigma_1,\sigma_2],~ r_i \leq \Delta_2^i:= \bm{1}^\top [\bm{f}_{i}(\bm{1}\pi^+)]^{\overline{\bm{d}}_{i}}_{\underline{\bm{d}}_{i}}$
        \item $r_\Nc\in [\sigma_3,\sigma_4],~ r_i \geq \Delta_3^i:= \bm{1}^\top [\bm{f}_{i}(\bm{1}\pi^-)]^{\overline{\bm{d}}_{i}}_{\underline{\bm{d}}_{i}}$
        \item $r_\Nc\in [\sigma_2,\sigma_3],~ \bm{1}^\top \bm{d}_i^{\psi^\ast}(\bm{1}\pi^\ast)= r_i$,
    \end{enumerate}
    under which $\mbox{VoC}^{\chi^\ast,\mbox{\tiny NEM}}_i= 0$.
\end{corollary}
\noindent {\em Proof:} See Appendix \ref{sec:AppProofs}.\hfill$\Box$\\
Dually, Corollary \ref{corol:PositiveVoC} (depicted in Fig.\ref{fig:CorolDepiction}) gives the conditions for strict individual rationality. As shown in Fig.\ref{fig:CorolDepiction}, the corollary indicates that whenever the community OEs constraints are binding, all members achieve a strictly positive benefit over their benchmark. Also, if neither the community OEs nor the benchmark OEs are binding, whenever the benchmark's optimal net consumption sign $\sgn\{z^{\mbox{\tiny NEM}^\ast}_i(r_i)\}$ (derived in Lemma 2 in \cite{Alahmed&Cavraro&Bernstein&Tong:23AllertonArXiv}) opposes the community's aggregate net consumption sign $\sgn\{z^{\psi^\ast}_{\Nc}(r_{\Nc})\}$, \ie $\sgn\{z^{\mbox{\tiny NEM}^\ast}_i(r_i)\} \neq \sgn\{z^{\psi^\ast}_{\Nc}(r_{\Nc})\}$ the member is strictly better than its benchmark. The yellow point in Fig.\ref{fig:CorolDepiction} represents condition (iii) in Corollary \ref{corol:PositiveVoC}.

The following proposition shows the monotonicity of the VoC of every member with the OEs.

\begin{proposition}[VoC comparative statics analysis]\label{prop:VoCwithpi}
For every member $i \in \Nc$ under the OEs-aware pricing policy, the value of joining the community $\mbox{VoC}_i^{\chi^\ast,\mbox{\tiny NEM}}$ monotonically increases with the import OE $\overline{z}_{\Nc}$ and monotonically decreases with the export OE $\underline{z}_{\Nc}$.

Furthermore, $\mbox{VoC}_i^{\chi^\ast,\mbox{\tiny NEM}}$ monotonically increases with $\pi^+$, except $r_{\Nc}\leq \sigma_1, \forall r_i \in \mathbb{R}_+$ and when $r_{\Nc}\in (\sigma_1,\sigma_2), r_i \leq \Delta_1^i$, under which the $\mbox{VoC}_i^{\chi^\ast,\mbox{\tiny NEM}}$ monotonically decreases. 

Lastly, $\mbox{VoC}_i^{\chi^\ast,\mbox{\tiny NEM}}$ monotonically decreases with $\pi^-$, except when $r_{\Nc}\geq \sigma_4, \forall r_i \in \mathbb{R}_+$ and when $r_{\Nc}\in (\sigma_3,\sigma_4), r_i \leq \Delta_4^i$ under which the $\mbox{VoC}_i^{\chi^\ast,\mbox{\tiny NEM}}$ monotonically increases.
\end{proposition}
\noindent {\em Proof:} See Appendix \ref{sec:AppProofs}.\hfill$\Box$\\
Table \ref{tab:PropDepiction} summarizes the detailed comparative static analysis in Proposition \ref{prop:VoCwithpi} by considering $\epsilon$-increases of the exogenous parameters ($\pi^+,\pi^-,\overline{z}_{\Nc},\underline{z}_{\Nc}$). The monotonicity with ($\overline{z}_{\Nc},\underline{z}_{\Nc}$) indicates the increased willingness of DSO prosumers to join the community when the community's OEs become less restrictive than their aggregate individual OEs, which is a straightforward benefit of aggregation. From Table \ref{tab:PropDepiction}, increasing $\pi^+$ increases the VoC when the community price is $<\pi^+$ and when the community price is $\pi^+$, but the benchmark's import OE does not bind. Conversely, increasing $\pi^-$ decreases the VoC decreases whenever the community price is $>\pi^-$ and when the community price is $\pi^-$, but the benchmark's export OE does not bind.

\begin{table}[]
\centering
\caption{VoC Comparative Statics (Proposition \ref{prop:VoCwithpi}).}
\vspace{-0.25cm}
\label{tab:PropDepiction}
\resizebox{\columnwidth}{!}{%
\begin{tabular}{@{}cccccc@{}}
\toprule \midrule
                   -----                & $r_{\Nc} \leq \sigma_1$ & $r_{\Nc} \in (\sigma_1, \sigma_2)$ & $r_{\Nc} \in [\sigma_2, \sigma_3]$ & $r_{\Nc} \in (\sigma_3, \sigma_4)$ & $r_{\Nc} \geq \sigma_4$ \\ \midrule
$r_i \leq \Delta_1^i$              &      $\color{red} \pmb{\downarrow} $        $\color{blue} 0 $    $\color{teal} \pmb{\uparrow}  $    $\color{orange} 0  $    &       $\color{red} \pmb{\downarrow} $     $\color{blue} 0 $      $\color{teal} 0 $            $\color{orange} 0  $          &            $\color{red} \pmb{\uparrow} $       $\color{blue} 0 $     $\color{teal} 0 $        $\color{orange} 0  $        &        $\color{red} \pmb{\uparrow} $     $\color{blue} \pmb{\downarrow} $            $\color{teal} 0 $       $\color{orange} 0  $        &          $\color{red} \pmb{\uparrow} $          $\color{blue} \pmb{\uparrow} $    $\color{teal} 0 $   $\color{orange} \pmb{\downarrow}  $ \\
$r_i \in (\Delta_1^i,\Delta_2^i)$  &            $\color{red} \pmb{\downarrow} $      $\color{blue} 0 $    $\color{teal} \pmb{\uparrow}  $  $\color{orange} 0  $   &     $\color{red} 0 $         $\color{blue} 0 $            $\color{teal} 0 $       $\color{orange} 0  $       &                   $\color{red} \pmb{\uparrow} $               $\color{blue} 0 $   $\color{teal} 0 $   $\color{orange} 0  $ &                  $\color{red} \pmb{\uparrow} $    $\color{blue} \pmb{\downarrow} $       $\color{teal} 0 $       $\color{orange} 0  $    &         $\color{red} \pmb{\uparrow} $    $\color{blue} \pmb{\uparrow} $     $\color{teal} 0 $      $\color{orange} \pmb{\downarrow}  $  \\
$r_i \in [\Delta_2^i,\Delta_3^i]$  &          $\color{red} \pmb{\downarrow} $    $\color{blue} 0 $    $\color{teal} \pmb{\uparrow}  $    $\color{orange} 0  $     &     $\color{red} \pmb{\uparrow} $       $\color{blue} 0 $             $\color{teal} 0 $       $\color{orange} 0  $        &                     $\color{red} 0 $         $\color{blue} 0 $     $\color{teal} 0 $   $\color{orange} 0  $  &                 $\color{red} 0 $      $\color{blue} \pmb{\downarrow} $       $\color{teal} 0 $     $\color{orange} 0  $      &              $\color{red} 0 $   $\color{blue} \pmb{\uparrow} $     $\color{teal} 0 $      $\color{orange} \pmb{\downarrow}  $ \\
$r_i \in (\Delta_3^i, \Delta_4^i)$ &         $\color{red} \pmb{\downarrow} $        $\color{blue} \pmb{\downarrow} $    $\color{teal} \pmb{\uparrow}  $   $\color{orange} 0  $   &       $\color{red} \pmb{\uparrow} $     $\color{blue} \pmb{\downarrow} $          $\color{teal} 0 $          $\color{orange} 0  $        &                    $\color{red} 0 $       $\color{blue} \pmb{\downarrow} $     $\color{teal} 0 $      $\color{orange} 0  $  &                $\color{red} 0 $    $\color{blue} 0 $         $\color{teal} 0 $     $\color{orange} 0  $       &              $\color{red} 0 $    $\color{blue} \pmb{\uparrow} $   $\color{teal} 0 $     $\color{orange} \pmb{\downarrow}  $  \\
$r_i \geq \Delta_4^i$              &         $\color{red} \pmb{\downarrow} $    $\color{blue} \pmb{\downarrow} $     $\color{teal} \pmb{\uparrow}  $    $\color{orange} 0  $     &     $\color{red} \pmb{\uparrow} $      $\color{blue} \pmb{\downarrow} $          $\color{teal} 0 $            $\color{orange} 0  $       &                      $\color{red} 0 $        $\color{blue} \pmb{\downarrow} $     $\color{teal} 0 $    $\color{orange} 0  $  &                  $\color{red} 0 $    $\color{blue} \pmb{\uparrow} $       $\color{teal} 0 $       $\color{orange} 0  $     &             $\color{red} 0 $     $\color{blue} \pmb{\uparrow} $  $\color{teal} 0 $    $\color{orange} \pmb{\downarrow}  $    \\ \midrule \bottomrule
\end{tabular}%
} \vspace{0.05cm}
\begin{adjustwidth}{1pt}{1pt} \footnotesize{\begin{flushleft} $^\ddagger$Each color represents $\epsilon$-increases of one exogenous parameter ($\pi^+,\pi^-,\overline{z}_{\Nc},\underline{z}_{\Nc}$) while fixing all others.\\ \end{flushleft}\begin{center} $\color{red} \pi^+$: retail rate,~ $\color{blue} \pi^-$: export rate,~ $\color{teal} \overline{z}_{\Nc}$: import OE,~ $\color{orange} \underline{z}_{\Nc}$: export OE.\\ \end{center}
\begin{center}$\pmb{\uparrow}$: increasing,~ $\pmb{\downarrow}$: decreasing,~ $0$: unchanged. 
 \end{center}}
\end{adjustwidth}
\end{table}

Lastly, we draw a stronger statement about the optimality of member surplus under a community with aggregate-level OEs compared to a community with member-level OEs.

\begin{theorem}[Surplus optimality]\label{thm:AgglevelVSMemLevel}
Given $r_i$, and $r_\Nc$, if an aggregate-level OEs and a member-level OEs communities are both in the net-consuming or net-producing zones then the surplus of a prosumer in the former community is no less than its surplus in the latter, \ie if $r_{\Nc}< \sigma_2$ and $r_{\Nc} < \Theta_1(\bm{r})$ or $r_{\Nc}> \sigma_3$ and $r_{\Nc} > \Theta_2(\bm{r})$ then
$$S^{\chi^\ast}_i(\bm{\psi}^\ast_i(r_i),r_i) \geq S^{\ast,\mbox{\tiny DNEM}}_i(r_i), $$
where $S^{\ast,\mbox{\tiny DNEM}}_i(r_i)$ is the maximum member surplus under D-NEM with member-level OEs, and ($\Theta_1(\bm{r}), \Theta_2(\bm{r})$) are D-NEM thresholds \cite{Alahmed&Cavraro&Bernstein&Tong:23AllertonArXiv}. 
\end{theorem}
\noindent {\em Proof:} See Appendix \ref{sec:AppProofs}.\hfill$\Box$\\
Whereas the {\em welfare} sub-optimality of a community with member-level OEs to a community with aggregate-level OEs is straightforward, as the former has a smaller feasible solution space, the {\em surplus} sub-optimality of members in the member-level OEs community to members in the aggregate-level OEs is less obvious. Roughly speaking, the reason for this {\em surplus} sub-optimality is that enforcing OEs at the members' meters rather than at the community's revenue meter, restricted energy sharing. Indeed, when the member-level OEs are binding, members who were buyers under aggregate-level OEs are now forced to reduce their consumption, and members who were sellers are now forced to increase their self-consumption instead. This prevented both buyers and sellers from transacting at the community price, which is always better than the DSO's NEM price. A direct result of Lemma \ref{lem:CostCausation}, Theorem 1 in \cite{Alahmed&Cavraro&Bernstein&Tong:23AllertonArXiv}, and Theorem \ref{thm:AgglevelVSMemLevel} is the following surplus order
$$S^{\chi^\ast}_i(\bm{\psi}^\ast_i(r_i),r_i) \geq S^{\ast,\mbox{\tiny DNEM}}_i(r_i) \geq S^{\mbox{\tiny NEM}^\ast}_i(r_i),$$
where the first inequality holds only under the conditions in Theorem \ref{thm:AgglevelVSMemLevel}.

%% file: num.tex
We constructed hypothetical energy community with $N=20$ residential customers (4 of which do not have BTM generation), whereby the 20 households pool and aggregate their resources behind a DSO revenue meter under a NEM tariff. A one-year DER data\footnote{We used 2018 \href{https://www.pecanstreet.org/dataport/}{PecanStreet data} for households in Austin, TX with 15-minute granularity.} was used. The DSO's NEM tariff has a ToU-based ({\em retail rate} $\pi^+$) with $\pi^+_{\mbox{\tiny ON}}=\$0.40$/kWh and $\pi^+_{\mbox{\tiny OFF}}=\$0.20$/kWh as on- and off-peak prices, respectively, and an ({\em export rate} $\pi^-$) that is based on the wholesale market price.\footnote{We used the averaged 2018 real-time wholesale prices in Texas. The data is accessible at \href{https://www.ercot.com/mktinfo/prices}{ERCOT}.} The DSO OEs are varied, but we assumed $\sum_{i \in \mathcal{N}} \overline{z}_i = \overline{z}_\mathcal{N}=-\underline{z}_\mathcal{N}=-\sum_{i \in \mathcal{N}} \underline{z}_i$.

We assumed a quadratic concave and non-decreasing utility function that represents the satisfaction from total consumption $d_i:=\bm{1}^\top \bm{d}_i$ for every $i\in \Nc$, given by
\begin{equation}\label{eq:UtilityForm}
   U_{i}(d_{i})=\left\{\begin{array}{ll}
\alpha_{i} d_{i}-\frac{1}{2}\beta_{i} d_{i}^2,\hspace{-0.2cm} &\hspace{-0.2cm} 0 \leq d_{i} \leq \frac{\alpha_{i}}{\beta_{i}} \\
\frac{\alpha_{i}^2}{2 \beta_{i}},\hspace{-0.2cm} &\hspace{-0.2cm} d_{i}>\frac{\alpha_{i}}{\beta_{i}},
\end{array} \right.
\end{equation}
where $\alpha_{i}, \beta_{i}$ are parameters that were learned and calibrated using historical retail prices\footnote{We used \href{https://data.austintexas.gov/stories/s/EOA-C-5-a-Austin-Energy-average-annual-system-rate/t4es-hvsj/}{Data.AustinTexas.gov} historical residential rates in Austin, TX.} and consumption,\footnote{We used pre-2018 PecanStreet data for households in Austin, TX.} and by assuming an elasticity of 0.21 taken from \cite{ASADINEJAD_Elasticity:18EPSR} (see appendix B in \cite{Alahmed&Tong:22IEEETSG}).

\begin{comment}
\subsection{Community Welfare}
Figure \ref{fig:NormalizedW} shows the average monthly normalized welfare and aggregate net consumption of aggregate-level OEs (solid curves) and member-level OEs (dashed curves) communities (see Fig.\ref{fig:EnergyCommunity}) compared to the benchmark prosumers (dotted curve). The welfare of the community with aggregate-level OEs was consistently higher than the welfare of the community with member-level OEs. Increasing the OEs, increased the welfare of both community frameworks, and reduced the welfare gap between them, as the OEs were binding less often. At high enough OEs, the welfare gap between the two community frameworks eventually vanished.
The aggregate net-consumption curves show that the welfare optimality is tied to how much MWs are pooled and aggregated.

\begin{figure}
    \centering
    \includegraphics[scale=0.35]{figures/NormalizedWelfare.eps}
    \vspace{-0.3cm}
    \caption{Normalized welfare and aggregate net-consumption of aggregate-level OEs (solid curves) and member-level OEs (dashed curves) communities compared to the benchmark prosumers (dotted curves).}
    \label{fig:NormalizedW}
\end{figure}
\end{comment}

\subsection{Community Members' Payments and Surpluses}
\begin{figure*}
    \centering
    \includegraphics[scale=0.53]{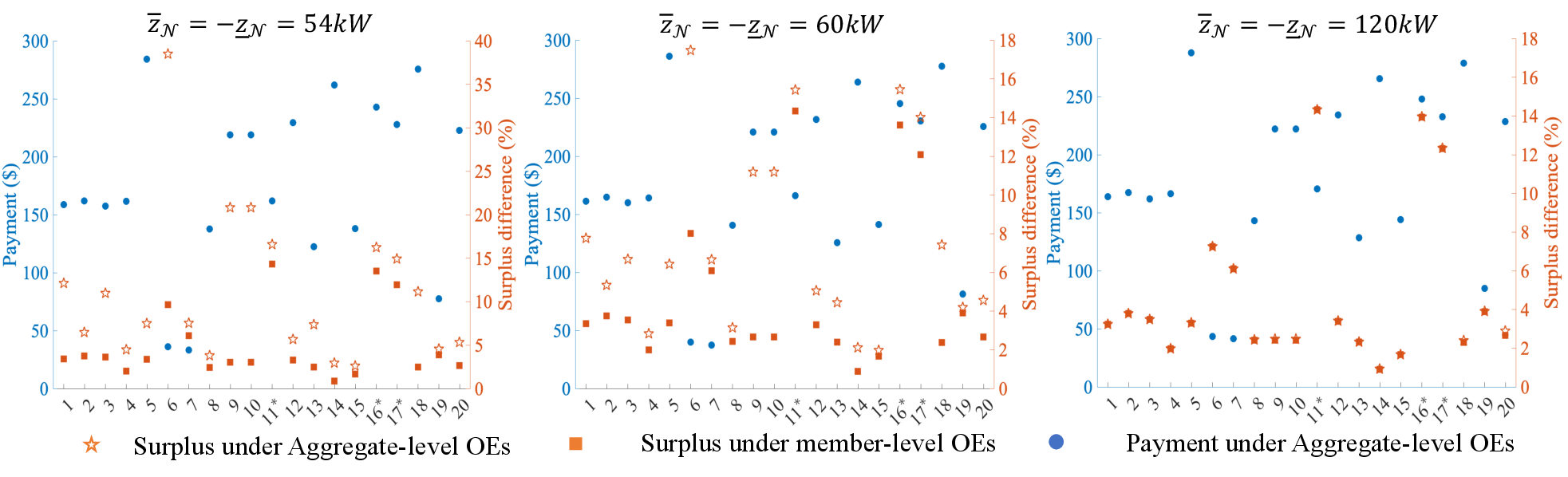}
    \vspace{-0.5cm}
    \caption{Percentage surplus difference and payment for the $N=20$ members. Members with black asterisks are those with no BTM generation.}
    \label{fig:SurNum}
\end{figure*}

Figure \ref{fig:SurNum} shows the average monthly community member's payment, percentage surplus difference over the benchmark, and percentage surplus difference over the benchmark when the community has member-level OEs.

Four main observations on the surplus difference (red asterisks and squares) are in order. First, members under communities with aggregate-level and member-level OEs achieved a higher surplus than their benchmark (Theorem \ref{thm:Equilibrium} and Theorem 1 in \cite{Alahmed&Cavraro&Bernstein&Tong:23AllertonArXiv}). For members who achieved higher surplus differences, it indicates that their benchmarks experienced high surplus loss due to standalone customer OEs. Second, tightening the OEs, made it, in general, more advantageous to join the community. The average surplus difference of the 20 customers under 120kW, 60kW, and 54kW OEs was approximately 5\%, 8\%, and 12\%, respectively. Third, community members under aggregate-level OEs achieved a higher surplus than community members under member-level OEs (Theorem \ref{thm:AgglevelVSMemLevel}), and the difference was higher when the OEs were tighter. The average difference between the gain of members under communities with aggregate-level and member-level OEs was 0.03\%, 2.91\%, and 6.14\% for 120kW, 60kW, and 54kW OEs, respectively. Fourth, non-solar adopters (members with asterisks) suffered the most as the OEs became tighter, which is because when they operate in standalone settings they were highly vulnerable to the import rate. One can see that although customers 11$^\ast$, 16$^\ast$, and 17$^\ast$ were the highest surplus difference achievers under 120kW OEs, their performance significantly degraded when the OEs were tightened to 54kW. 

\par Additionally, one can see that the average monthly payment (blue dots) of aggregate-level OEs framework's community members did not change much when the OEs were tightened, because the fixed reward compensated customers when they faced $>\pi^+$ buy rates (when the import OE was binding) and $<\pi^-$ sell rates (when the export OE was binding). The volumetric to fixed reward ratio (Table \ref{tab:TariffRatio}) helps in understanding this phenomenon. When OEs were relaxed (\ie in the 120kW case), the ratio was effectively $\infty$ as the OEs were never binding. As the OEs were further constrained to 60kW and 54kW, the average ratio reduced to 39.4\% and 19.55\%, respectively, which shows that fixed rewards increased to compensate for the loss in volumetric charge, keeping the overall payment only slightly changed.

\begin{table*}[]
\centering
\caption{Absolute value of the percentage ratio of volumetric to fixed charge (\%).}
\label{tab:TariffRatio}
\vspace{-0.3cm}
\begin{tabular}{ccccccccccccccccccccc}\toprule \midrule
$\overline{z}_\Nc = -\underline{z}_\Nc$ & 1 & 2 & 3 & 4 & 5 & 6 & 7 & 8 & 9 & 10 & 11$^\ast$ & 12 & 13 & 14 & 15 & 16$^\ast$ & 17$^\ast$ & 18 & 19 & 20 \\ \midrule
54kW & 17 & 17 & 17 & 17 & 30 & 5 & 4 & 15 & 23 & 23 & 17 & 24 & 13 & 27 & 15 & 25 & 24 & 29 & 9 & 23 \\
60kW & 34 & 34 & 33 & 34 & 59 & 9 & 9 & 29 & 45 & 45 & 34 & 48 & 26 & 54 & 29 & 50 & 47 & 57 & 17 & 46 \\
120kW & \multicolumn{20}{c}{ zero fixed charge/reward (One-part pricing)}\\ \midrule \bottomrule
\end{tabular} 
\end{table*}

\subsection{Daily Community Price}
\begin{figure}
    \centering
    \includegraphics[scale=0.38]{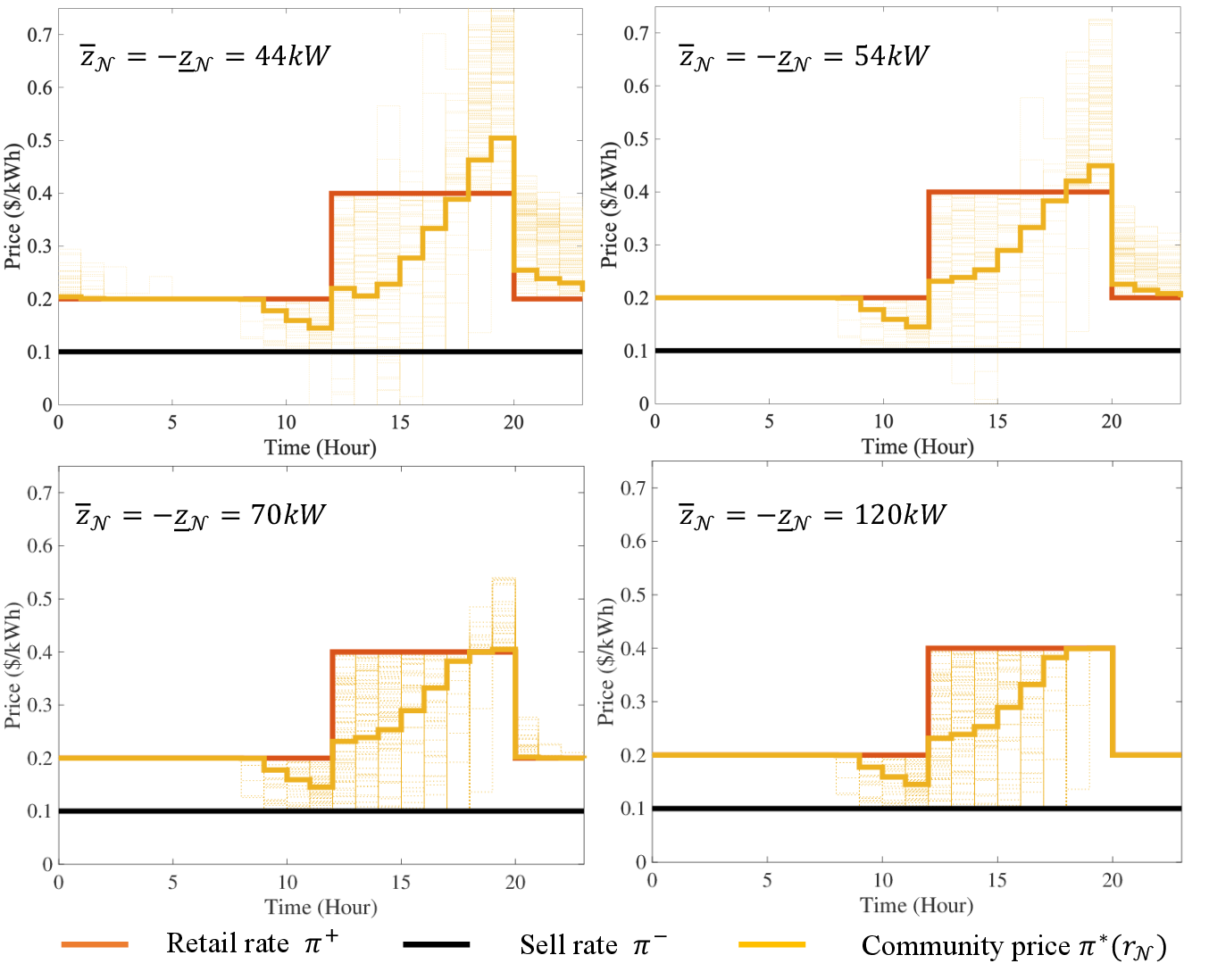}
    \vspace{-0.40cm}
    \caption{Average community price under different OEs.}
    \label{fig:OEsPrice}
\end{figure}

Figure \ref{fig:OEsPrice} shows the aggregate-level OEs community's daily price over a year (dotted curves) and the average daily price (yellow curves) under different OEs. For this particular figure, we set the sell rate at $\pi^-=\$0.1$/kWh. In hours with binding import OE, the community price became higher than $\pi^+$, as the operator charges a higher price to reduce consumption. Whereas, when the export OE was binding, the price became less than $\pi^-$, as the operator charges a lower price to encourage higher consumption, hence locally consuming the excess renewables. Therefore, when the OEs were set at 44kW, the community price exceeded the NEM X rates in many instances, as the OEs were binding more often. The same goes for 54kW OEs. By further relaxing the OEs, the community price stayed within the NEM X rates, as it dynamically ranged between $\pi^+$ and $\pi^-$ passing by $\chi^z$. Indeed, at 120kW OEs, the daily price never oscillated between and never exceeded $\pi^+$ and $\pi^-$, mimicking the community pricing policy without OEs, \ie Dynamic NEM in \cite{Alahmed&Tong:24TEMPR}.

\subsection{Community Price and Aggregate Net Consumption}
\begin{figure}
    \centering
    \includegraphics[scale=0.51]{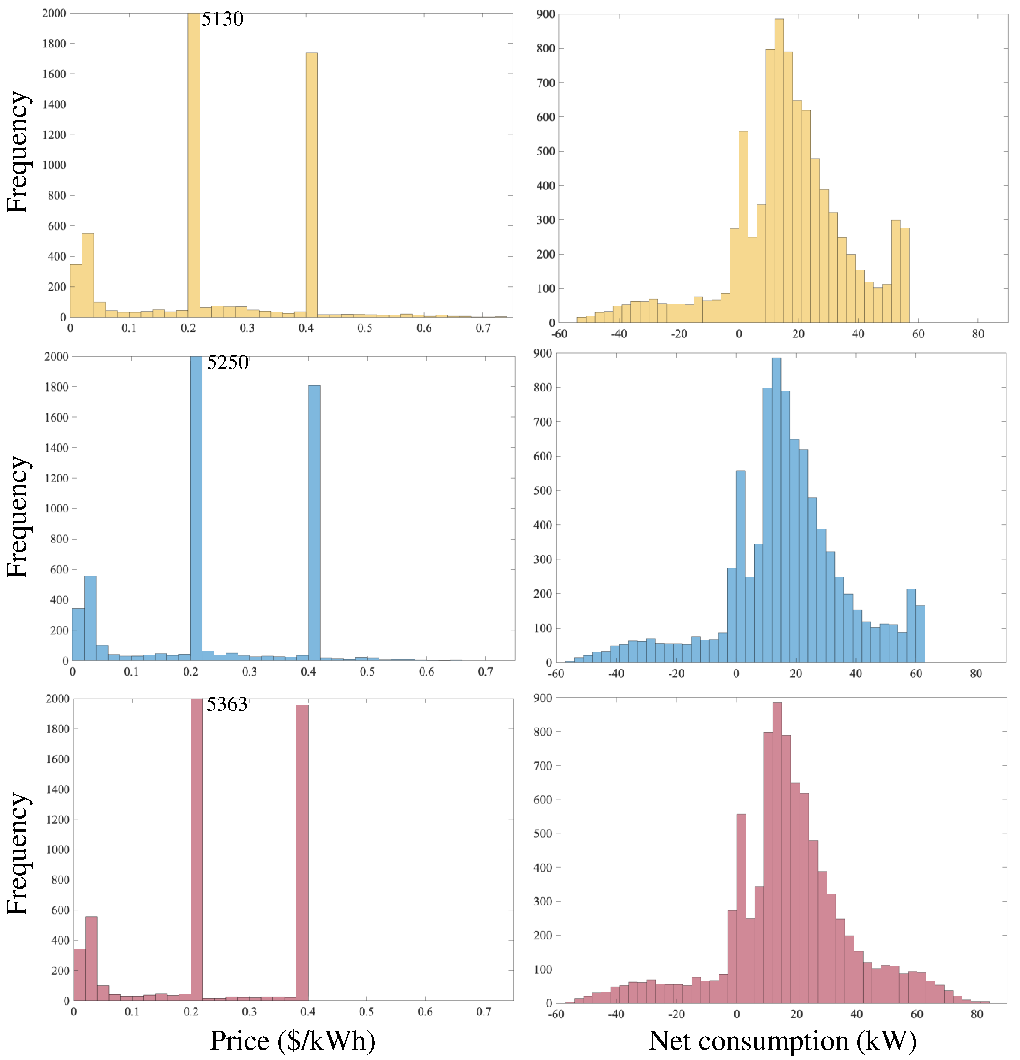}
    \vspace{-0.7cm}
    \caption{Histogram of hourly community price (left column) and aggregate net-consumption (right column) under OEs $\overline{z}_\mathcal{N}=-\underline{z}_\mathcal{N}=54$kW (top), $\overline{z}_\mathcal{N}=-\underline{z}_\mathcal{N}=60$kW (middle), and $\overline{z}_\mathcal{N}=-\underline{z}_\mathcal{N}=120$kW (bottom). }
    \label{fig:GamNum}
\end{figure}

Figure \ref{fig:GamNum} shows histograms of the hourly aggregate-level OEs community price (left column) and aggregate net consumption (right column), over the studied year, under OEs $\overline{z}_\mathcal{N}=-\underline{z}_\mathcal{N}=54$kW (top), $\overline{z}_\mathcal{N}=-\underline{z}_\mathcal{N}=60$kW (middle), and $\overline{z}_\mathcal{N}=-\underline{z}_\mathcal{N}=120$kW (bottom). First, note that the frequency of the prices \$0.4/kWh and \$0.2/kWh was high because the ToU retail rate $\pi^+$ can only take these values, whereas the sell rate $\pi^-$ can take any value between \$0.04/kWh to \$0.15/kWh. When the OEs were high, $\overline{z}_\mathcal{N}=-\underline{z}_\mathcal{N}=120$kW, the community price never exceeded $\pi^+$, as the import OE was never binding. When the OE decreased to $\overline{z}_\mathcal{N}=-\underline{z}_\mathcal{N}=60$kW, the frequency of \$0.4/kWh and \$0.2/kWh decreased as the community price crossed $\pi^+$ (note the higher frequency of prices $\chi^+_{\mbox{\tiny ON}}>\$0.4$/kWh and $\chi^+_{\mbox{\tiny OFF}}>\$0.2$/kWh) whenever the import rate was binding. This phenomenon was even clearer when the OEs were decreased to $\overline{z}_\mathcal{N}=-\underline{z}_\mathcal{N}=54$kW, under which $\chi^+_{\mbox{\tiny ON}}$ exceeded \$0.7/kWh in some hours. The same observation can be extended to the export side, under which the community is charged with $\chi^- <\pi^-$, whenever the export OE was binding. This however was not clear from the plots, because the community's generation was relatively low compared to consumption.

The dynamic aggregate net consumption plots (right column) show how the community price managed the community's operation at the PCC, by inducing its members to consume more when the export OE was binding and consume less when the import OE was binding.  

%% file: appendixA/appendixA.tex
From \cite{Alahmed&Cavraro&Bernstein&Tong:23AllertonArXiv}, if the community has member-level OEs instead of aggregate-level OEs, the community pricing that induces market equilibrium is the D-NEM pricing given below.
\begin{policyy*}
For every $i\in \Nc$, the threshold-based and OEs-aware energy community pricing policy is 
$$\chi^{\ast}: \bm{r} \mapsto P^{\mbox{\tiny DNEM}}_i(z_i)=\pi^{\mbox{\tiny DNEM}}(\bm{r})\cdot z_i,$$ with
\begin{align}\label{eq:MemberLevelPricingMechanism}
 \pi^{\mbox{\tiny DNEM}}(\bm{r}) &= \begin{cases}
 \pi^+ & ,     r_{\Nc}< \Theta_1(\bm{r})  \\ 
\pi^z(\bm{r}) & , r_{\Nc}\in [\Theta_1(\bm{r}),\Theta_2(\bm{r})]\\ 
\pi^- & , r_{\Nc} > \Theta_2(\bm{r}),
\end{cases}
 \end{align}
 where the thresholds $\Theta_2(\bm{r}) \geq \Theta_1(\bm{r})$ are computed as
\begin{align}
\Theta_1(\bm{r}):=\sum_{i=1}^N [D_i^{+}]^{\overline{z}_i+r_i}_{\underline{z}_i+r_i},~~
\Theta_2(\bm{r}):=\sum_{i=1}^N [D_i^{-}]^{\overline{z}_i+r_i}_{\underline{z}_i+r_i},\nn
\end{align}
with $
 D_i^{+}:= [f_i\left( \pi^{+}\right)]^{\overline{d}_i}_{0},~
 D_i^{-} := [f_i\left( \pi^{-}\right)]^{\overline{d}_i}_{0}$,
and the $\max$ and $\min$ operators being elementwise.

The price $\pi^z(\bm{r}):=\mu^\ast(\bm{r})\in [\pi^-,\pi^+]$ is the solution of
\begin{equation}
    \sum_{i=1}^N [D_i^{z}(\mu)]^{\overline{z}_i+r_i}_{\underline{z}_i+r_i}= r_{\Nc},\nn
\end{equation}
where $D_i^{z}(\mu):=  [f_i\left( \mu(\bm{r})\right)]^{\overline{d}_i}_{0}$. 
\end{policyy*}

%% file: appendixB_v2/BenchmarkLem_v2.tex
Lemma \ref{lem:BenchmarkSur} (Lemma 2 in \cite{Alahmed&Cavraro&Bernstein&Tong:23AllertonArXiv}) characterizes the optimal response and maximum surplus of the benchmark prosumer.
\begin{lemma}[Benchmark prosumer optimal response and maximum surplus]\label{lem:BenchmarkSur}
The optimal benchmark consumption of every member $i\in \Nc$ under the DSO's NEM regime abides by the following 4-thresholds,
\begin{align*}
 \Delta_1^i &:= \Delta_2^i - \overline{z}_i, \quad \quad\quad \quad\quad  \Delta_2^i := [f_{i}(\pi^+)]^{\overline{d}_{i}}_{0}\nn\\
\Delta_3^i &:=  [f_{i}(\pi^-)]^{\overline{d}_{i}}_{0}, \quad \quad \quad \quad
\Delta_4^i := \Delta_3^i - \underline{z}_i,\nn
\end{align*}
as 
\begin{align}
    d^{\mbox{\tiny NEM}^\ast}_i(r_i) &=\begin{cases}
d^{\mu^+}_i(r_i) &, r_i\leq \Delta_1^{i}\\ 
d^{\pi^+}_i &, r_i \in (\Delta_1^{i},\Delta_2^{i})\\ 
d^{\mu^o}_i(r_i) &, r_i \in [\Delta_2^{i},\Delta_3^{i}]\\  
d^{\pi^-}_i &, r_i \in (\Delta_3^{i},\Delta_4^{i})\\ 
d^{\mu^-}_i(r_i) &, r_i \geq \Delta_4^{i},
\end{cases}\label{eq:OptConsStandalone}\\
z^{\mbox{\tiny NEM}^\ast}_i(r_i) &= d^{\mbox{\tiny NEM}^\ast}_i(r_i) - r_i,
\end{align}
where the static consumptions $d^{\pi^+}_{i}$ and $d^{\pi^-}_{i}$ and dynamic consumptions $d^{\mu^+}_{i}(r_i),d^{\mu^o}_i(r_i),d^{\mu^-}_i(r_i)$, for every $k \in \mathcal{K}$, are all given by $ d^x_{ik} := [f_{i}(x)]^{\overline{d}_{i}}_{0}$,
 and the prices $\mu^+(r_i), \mu^o(r_i), \mu^-(r_i)$ are, respectively, the solutions of:
 \begin{align*}
     [f_{i}(\mu^+)]^{\overline{d}_{i}}_{0} = \overline{z}_i + r_i,~
     [f_{i}(\mu^{o})]^{\overline{d}_{i}}_{0} =  r_i,~
     [f_{i}(\mu^-)]^{\overline{d}_{i}}_{0} = \underline{z}_i + r_i,\nn
 \end{align*}
 with the order $\mu^+(r_i)\geq \pi^+\geq \mu^o(r_i)\geq \pi^-\geq \mu^-(r_i)$.

 Therefore, by definition, the surplus under optimal consumption is given by:
 \begin{equation}\label{eq:OptSStandalone}
    S^{\mbox{\tiny NEM}^\ast}_i(r_i) :=\begin{cases}
U_i(d^{\mu^+}_i(r_i)) - \pi^+ \overline{z}_i &, r_i\leq \Delta_1^{i}\\ 
U_i(d^{\pi^+}_i) - \pi^+ (d^{\pi^+}_i- r_i) &, r_i \in (\Delta_1^{i},\Delta_2^{i})\\ 
U_i(d^{\mu^o}_i(r_i)) &, r_i \in [\Delta_2^{i},\Delta_3^{i}]\\  
U_i(d^{\pi^-}_i) - \pi^- (d^{\pi^-}_i- r_i) &, r_i \in (\Delta_3^{i},\Delta_4^{i})\\ 
U_i(d^{\mu^-}_i(r_i)) - \pi^- \underline{z}_i &, r_i \geq \Delta_4^{i}.
\end{cases}
\end{equation}

\end{lemma}

%% file: appendixB_v2/appendixB_v2.tex
\subsection{Proof of Lemma \ref{lem:OptSchedule}}
\input{appendixB_v2/MemberLem_v2}

\subsection{Proof of Lemma \ref{lem:CostCausation}}
\input{appendixB_v2/CostCausation_v2}

\subsection{Lemma \ref{lem:CentralizedWelfare} and its proof}
\input{appendixB_v2/WelfareLem_v2}

\subsection{Proof of Theorem \ref{thm:Equilibrium}}
\input{appendixB_v2/Equilibrium_v2}

\subsection{Proof of Proposition \ref{corol:PriceOrder}}
\input{appendixB_v2/PriceOrderLem}

\subsection{Proof of Corollary \ref{corol:PositiveVoC}}
\input{appendixB_v2/VoCPve}
\subsection{Proof of Proposition \ref{prop:VoCwithpi}}
\input{appendixB_v2/CompStatics}

\subsection{Proof of Theorem \ref{thm:AgglevelVSMemLevel}}
\input{appendixB_v2/SurplusLem}

%% file: appendixB_v2/MemberLem_v2.tex
Recall the community member's surplus-maximization problem under the proposed market mechanism (\ref{eq:argmax_d+}). One can easily see that the objective is both concave and differentiable. From KKT conditions, the optimal consumption of every $i\in \Nc$ community member is:
$$d^{\psi^\ast}_{i}(\pi^\ast) = \max\{0, \min\{f_{i}(\pi^\ast),\overline{d}_{i}\}\}=: [f_{i}(\pi^\ast)]_{0}^{\overline{d}_{i}},$$
which is, given the OEs-aware pricing policy,
\begin{equation}\label{eq:OptdiMember}
       d^{\psi^\ast}_i(\pi^\ast) = \begin{cases}
           d_i^{\psi^\ast}(\chi^+(r_{\Nc})) &,  r_{\Nc} \leq \sigma_1\\
          d_i^{\psi^\ast}(\pi^+)&,  r_{\Nc} \in (\sigma_1,\sigma_2)\\
           d_i^{\psi^\ast}(\chi^z(r_{\Nc}))&,  r_{\Nc} \in [\sigma_2,\sigma_3]\\
            d_i^{\psi^\ast}(\pi^-)&,  r_{\Nc} \in (\sigma_3,\sigma_4)\\
            d_i^{\psi^\ast}(\chi^-(r_{\Nc}))&,  r_{\Nc} \geq \sigma_4.
       \end{cases}
   \end{equation}
By definition, the optimal net consumption is $z^{\psi^\ast}_i(\pi^\ast) =  d^{\psi^\ast}_i(\pi^\ast) - r_i$.
\hfill$\Box$

%% file: appendixB_v2/CostCausation_v2.tex
To prove Lemma \ref{lem:CostCausation}, we need to show that the OEs-aware pricing policy satisfies each of Axiom \ref{ax:equity}--Axiom \ref{ax:ProfitNeutrality}.

\subsubsection*{Axiom \ref{ax:equity}}
Since the OEs-aware pricing policy's price $\pi^\ast(r_{\Nc})$ in (\ref{eq:PricingMechanism}) is uniform, and the volumetric charge is of the form $\tilde{P}_i^{\chi^\ast}(z_i) = \pi^\ast z_i, \forall i\in \Nc$, it holds that for any two members $i,j \in \Nc, i \neq j$ if $z_i=z_j$, then
$$\tilde{P}_i^{\chi^\ast}(z_i) = \pi^\ast \cdot z_i = \pi^\ast \cdot z_j = \tilde{P}_j^{\chi^\ast}(z_j).$$

\subsubsection*{Axiom \ref{ax:monotonicity}}
For any member $i\in \Nc$, the payment under the OEs-aware pricing policy is monotonic in $z_i$ because
$\frac{\partial \tilde{P}_i^{\chi^\ast}}{\partial z_i}=\pi^\ast \geq 0.$
Additionally, the volumetric charge of every $i\in \Nc$ satisfies
$\tilde{P}_i^{\chi^\ast}(0) = \pi^\ast \cdot 0 = 0.$

\subsubsection*{Axiom \ref{ax:ProfitNeutrality}}
From Lemma \ref{lem:OptSchedule}, given the members' optimal consumption policy $\psi^\ast$ in response to the OEs-aware pricing policy $\chi^\ast$, the aggregate net consumption of the community $z^{\psi^\ast}_{\Nc}$ is as shown in (\ref{eq:AggOptz}), therefore,
 \begin{align*}
   &\Delta P^{\chi^\ast, \mbox{\tiny NEM}}(\bm{z}^{\psi^\ast}) := \sum_{i\in \Nc}P^{\chi^\ast}_i(z^{\psi^\ast}_i) -  P^{\mbox{\tiny NEM}}(z^{\psi^\ast}_\Nc)\\&=
\sum_{i\in \Nc} (\pi^\ast(r_{\Nc}) \cdot z^{\psi^\ast}_i(\pi^\ast) - A_i^\ast(r_{\Nc})) - \pi^{\mbox{\tiny NEM}}(z^{\psi^\ast}_\Nc) \cdot z^{\psi^\ast}_{\Nc}(\pi^\ast)\\&=
   \pi^\ast(r_{\Nc}) \cdot z^{\psi^\ast}_{\Nc}(\pi^\ast) - \sum_{i\in \Nc} A_i^\ast(r_{\Nc}) - \pi^{\mbox{\tiny NEM}}(z^{\psi^\ast}_\Nc) \cdot z^{\psi^\ast}_{\Nc}(\pi^\ast)
   \\&= (\pi^\ast(r_{\Nc}) - \pi^{\mbox{\tiny NEM}}(z^{\psi^\ast}_\Nc))\cdot z^{\psi^\ast}_{\Nc}(\pi^\ast) - \sum_{i\in \Nc} A_i^\ast(r_{\Nc})\\&\stackrel{\text{(\ref{eq:Pcommunity}),(\ref{eq:PricingMechanism}),(\ref{eq:AggOptz})}}{=} \begin{cases}
           (\chi^+(r_{\Nc})-\pi^+)\cdot\overline{z}_\Nc &\hspace{-0.2cm},  r_{\Nc} \leq \sigma_1\\
          (\pi^+-\pi^+)\cdot (d^{\psi^\ast}_\Nc(\pi^+) - r_{\Nc})&\hspace{-0.2cm},  r_{\Nc} \in (\sigma_1,\sigma_2)\\
           (\chi^z(r_{\Nc})-\pi^+)\cdot 0&\hspace{-0.2cm},  r_{\Nc} \in [\sigma_2,\sigma_3]\\
            (\pi^--\pi^-)\cdot(d^{\psi^\ast}_\Nc(\pi^-) - r_{\Nc})&\hspace{-0.2cm},  r_{\Nc} \in (\sigma_3,\sigma_4)\\
            (\chi^-(r_{\Nc})-\pi^-)\cdot \underline{z}_\Nc&\hspace{-0.2cm},  r_{\Nc} \geq \sigma_4.
       \end{cases}\\&~~~\quad - \sum_{i\in \Nc} \begin{cases}
(\chi^+(r_{\Nc})-\pi^+) \left(\overline{z}_i + \frac{\overline{z}_\Nc-\sum_{i\in \Nc}\overline{z}_i}{N}\right) &\hspace{-0.33cm} ,r_{\Nc}\leq \sigma_1 \\ 
(\chi^-(r_{\Nc})-\pi^-) \left(\underline{z}_i + \frac{\underline{z}_\Nc-\sum_{i\in \Nc}\underline{z}_i}{N}\right) &\hspace{-0.33cm}, r_{\Nc}\geq \sigma_4\\
0&\hspace{-0.43cm}, \text{otherwise},
\end{cases}\\
     &= \begin{cases}
           (\chi^+(r_{\Nc})-\pi^+)\cdot\overline{z}_\Nc &\hspace{-0.2cm},  r_{\Nc} \leq \sigma_1\\
          0 &\hspace{-0.2cm},  r_{\Nc} \in (\sigma_1,\sigma_2)\\
           0 &\hspace{-0.2cm},  r_{\Nc} \in [\sigma_2,\sigma_3]\\
            0 &\hspace{-0.2cm},  r_{\Nc} \in (\sigma_3,\sigma_4)\\
            (\chi^-(r_{\Nc})-\pi^-)\cdot \underline{z}_\Nc&\hspace{-0.2cm},  r_{\Nc} \geq \sigma_4.
       \end{cases} \\&~~ - \begin{cases}
(\chi^+(r_{\Nc})-\pi^+) \overline{z}_\Nc &\hspace{-0.27cm} ,r_{\Nc}\leq \sigma_1 \\ 
(\chi^-(r_{\Nc})-\pi^-) \underline{z}_\Nc &\hspace{-0.27cm}, r_{\Nc}\geq \sigma_4\\
0&\hspace{-0.30cm}, \text{otherwise}.
\end{cases}\\&= 0
 \end{align*}
 \subsubsection*{Axiom \ref{ax:rationality}}
\input{appendixB/IndRat}

By proving that the OEs-aware pricing policy satisfies Axiom \ref{ax:equity}--Axiom \ref{ax:ProfitNeutrality}, we showed that the OEs-aware pricing policy conforms with the cost-causation principle. 
\hfill$\Box$

%% file: appendixB/IndRat.tex
Given $r_i$, we compare the maximum benchmark surplus ($S^{\mbox{\tiny NEM}^\ast}_i(r_i)$) in (\ref{eq:OptSStandalone}) in Lemma \ref{lem:BenchmarkSur} and the community member surplus under the OEs-aware pricing policy, given by, using Lemma \ref{lem:OptSchedule},
\begin{align*}
    S^{\chi^\ast}_i(\psi^\ast_i(r_i),r_i)\hspace{-0.1cm} =  U_i(d_i^{\psi^\ast}(\pi^\ast)) - \pi^\ast ( d_i^{\psi^\ast}(\pi^\ast) - r_i) + A_i^\ast, 
\end{align*}
to show that individual rationality holds, \ie $\Delta S_i:= S_i^{\chi^\ast}(\psi^\ast_i(r_i),r_i) - S^{\mbox{\tiny NEM}^\ast}_i(r_i) \geq 0, \forall i \in \Nc$. For brevity, we assume non-binding consumption upper and lower limits. For notational convenience we define the following
\begin{align}\label{eq:ais}
    \overline{a}_i :=\hspace{-0.1cm} \left(\overline{z}_i + \frac{\overline{z}_\Nc-\sum_{i\in \Nc}\overline{z}_i}{N}\right)\hspace{-0.1cm},
    \underline{a}_i:= \hspace{-0.1cm}\left(\underline{z}_i + \frac{\underline{z}_\Nc-\sum_{i\in \Nc}\underline{z}_i}{N}\right),
\end{align}
hence the fixed reward can be re-written as
\begin{equation*}
     A^\ast_i(r_\Nc)= \begin{cases}
(\chi^+(r_\Nc)-\pi^+)\cdot \overline{a}_i, &\hspace{0cm} r_\Nc\leq \sigma_1 \\ 
(\chi^-(r_\Nc)-\pi^-)\cdot \underline{a}_i, &\hspace{0cm} r_\Nc\geq \sigma_4\\
0,&\hspace{0cm} \text{otherwise}.
\end{cases}
 \end{equation*}
Given that each of the piecewise functions $S^{\mbox{\tiny NEM}^\ast}_i$ and $S_i^{\chi^\ast}$ have 5 pieces, their difference $\Delta S_i$ have $5^2=25$ pieces. The proof is completed if we show that each piece in $\Delta S_i$ is non-negative. In \cite{Alahmed&Tong:24TEMPR}, we already showed the non-negativity of 9 of the 25 combinations, which are
  \begin{align*}
   &1)~r_\Nc\in (\sigma_1,\sigma_2), r_i \in (\Delta_1^i,\Delta_2^i)\\&2)~ r_\Nc\in (\sigma_1,\sigma_2), r_i \in [\Delta_2^i,\Delta_3^i]\hspace{4.2cm}\\
   &3)~r_\Nc\in (\sigma_1,\sigma_2), r_i \in (\Delta_3^i,\Delta_4^i)\\&4)~r_\Nc\in [\sigma_2,\sigma_3], r_i \in (\Delta_1^i,\Delta_2^i)\\&5)~r_\Nc\in [\sigma_2,\sigma_3], r_i \in [\Delta_2^i,\Delta_3^i]\\&6)~r_\Nc\in [\sigma_2,\sigma_3], r_i \in (\Delta_3^i,\Delta_4^i)\\ &7)~r_\Nc\in (\sigma_3,\sigma_4), r_i \in (\Delta_1^i,\Delta_2^i)\\ &8)~r_\Nc\in (\sigma_3,\sigma_4), r_i \in [\Delta_2^i,\Delta_3^i]\\
   &9)~r_\Nc\in (\sigma_3,\sigma_4), r_i \in (\Delta_3^i,\Delta_4^i).
  \end{align*}
To this end, we prove the non-negativity of the remaining 16 cases using the following property for concave and continuously differentiable function
$$L(x)(y-x)\geq U(y)-U(x)\geq L(y)(y-x),$$
where $x<y, \forall x,y\in \mathbb{R}$, and the community price order
$\chi^+(r_\Nc)\geq \pi^+ \geq \chi^z (r_\Nc) \geq \pi^- \geq \chi^-(r_\Nc)\geq 0.$, and the assumption $\sum_{i\in \Nc} \overline{z}_i \leq \overline{z}_\Nc,~~ \sum_{i\in \Nc} \underline{z}_i \geq \underline{z}_\Nc$

\begin{description}[align=left,leftmargin = 0pt]
  \item [(Case 1: $r_\Nc \leq \sigma_1, r_i \leq \Delta_1^i$)] The surplus difference is
  \begin{align*}
   \Delta S_i&=U_i(d^{\psi^\ast}_i(\chi^+)) - \chi^+ ( d^{\psi^\ast}_i(\chi^+)- r_i) +(\chi^+-\pi^+) \overline{a}_i \\&~~- U_i(d^{\mu^+}_i(r_i))+ \pi^+ \overline{z}_i.
    \end{align*}
    Here we have three sub-cases i) $d^{\psi^\ast}_i(\chi^+)> d^{\mu^+}_i$, ii) $d^{\psi^\ast}_i(\chi^+)< d^{\mu^+}_i$, and iii) $d^{\psi^\ast}_i(\chi^+)= d^{\mu^+}_i$. Under sub-case (i), we can write
    \begin{align*}
    \mu^+ (d^{\psi^\ast}_i(\chi^+)-d^{\mu^+}_i) &\geq U_i(d^{\psi^\ast}_i(\chi^+)) - U_i(d^{\mu^+}_i)\\&\geq \chi^+ (d^{\psi^\ast}_i(\chi^+)-d^{\mu^+}_i),
    \end{align*}
    where we used $L_i(d^{\psi^\ast}_i(\chi^+))= \chi^+$ and $L_i(d^{\mu^+})= \mu^+$. Using the lower bound above, we have
    $$\Delta S_i\geq (\chi^+-\pi^+)(\overline{a}_i-\overline{z}_i)\geq 0,$$
    where we used $\overline{z}_i= d^{\mu^+}_i - r_i$ when $r_i \leq \Delta_1^i$ (Lemma \ref{lem:BenchmarkSur}).\\
    Under sub-case (ii), we have
    \begin{align*}
    \chi^+ (d^{\mu^+}_i-d^{\psi^\ast}_i(\chi^+)) &\geq  U_i(d^{\mu^+}_i) - U_i(d^{\psi^\ast}_i(\chi^+))\\&\geq \mu^+ (d^{\mu^+}_i-d^{\psi^\ast}_i(\chi^+)).
    \end{align*}
    By multiplying the inequalities above by $-1$, we get the inequality in sub-case (i) and the proof follows directly.\\
    Under sub-case (iii), we have $\overline{z}_i = d^{\mu^+}_i - r_i = d^{\psi^\ast}_i(\chi^+)- r_i$, hence
    $$\Delta S_i\geq (\chi^+ - \pi^+) (\overline{a}_i-\overline{z}_i)\geq 0.$$
    
  \item [(Case 2: $r_\Nc \leq \sigma_1, \Delta_1^i < r_i < \Delta_2^i$)]  The surplus difference is
  \begin{align*}
   \Delta S_i&=U_i(d^{\psi^\ast}_i(\chi^+)) - \chi^+ ( d^{\psi^\ast}_i(\chi^+)- r_i) +(\chi^+-\pi^+) \overline{a}_i \\&~~- U_i(d^{\pi^+}_i)+ \pi^+ (d_i^{\pi^+}-r_i).
    \end{align*}
    Since $d^{\psi^\ast}_i(\chi^+) < d^{\pi^+}_i$, we have
    \begin{align*}
    \pi^+ (d^{\psi^\ast}_i(\chi^+)-d^{\pi^+}_i) &\geq U_i(d^{\psi^\ast}_i(\chi^+)) - U_i(d^{\pi^+}_i)\\&\geq \chi^+ (d^{\psi^\ast}_i(\chi^+)-d^{\pi^+}_i),
    \end{align*}
    and using the lower bound, we get
    $$\Delta S_i\geq (\chi^+ - \pi^+) (r_i - d^{\pi^+}_i + \overline{a}_i).$$
    We know that $\chi^+ \geq \pi^+$, so we only need to show that $r_i - d^{\pi^+}_i + \overline{a}_i\geq 0$.
    Note that by adding $\overline{a}_i$ to $d^{\pi^+}_i - \overline{z}_i < r_i < d^{\pi^+}_i$ (from case 2), we get
    $$0 \leq \overline{a}_i - \overline{z}_i < r_i - d^{\pi^+}_i + \overline{a}_i < \overline{a}_i,$$
    which proves that $\Delta S_i\geq (\chi^+ - \pi^+) (r_i - d^{\pi^+}_i + \overline{a}_i)\geq 0$.
    
  \item [(Case 3: $r_\Nc \leq \sigma_1, \Delta_2^i \leq r_i \leq \Delta_3^i$)] The surplus difference is
  \begin{align*}
   \Delta S_i&=U_i(d^{\psi^\ast}_i(\chi^+)) - \chi^+ ( d^{\psi^\ast}_i(\chi^+)- r_i) +(\chi^+-\pi^+) \overline{a}_i \\&~~- U_i(d^{\mu^o}_i).
    \end{align*}
    Since $d^{\psi^\ast}_i(\chi^+) < d^{\mu^o}_i$, we have
  \begin{align*}
  \mu^o (d^{\psi^\ast}_i(\chi^+)-d^{\mu^o}_i) &\geq U_i(d^{\psi^\ast}_i(\chi^+)) - U_i(d^{\mu^o}_i)\\&\geq \chi^+ (d^{\psi^\ast}_i(\chi^+)-d^{\mu^o}_i),
  \end{align*}
    and using the lower bound, we get
     $$\Delta S_i \geq \chi^+ (r_i - d^{\mu^o}_i) +(\chi^+ - \pi^+) \overline{a}_i,$$
     but we know that in Case 3, $r_i = d^{\mu^o}_i$, hence 
     $$\Delta S_i \geq (\chi^+ - \pi^+) \overline{a}_i\geq 0.$$
  \item [(Case 4: $r_\Nc \leq \sigma_1, \Delta_3^i < r_i < \Delta_4^i$)]  The surplus difference is
  \begin{align*}
   \Delta S_i&=U_i(d^{\psi^\ast}_i(\chi^+)) - \chi^+ ( d^{\psi^\ast}_i(\chi^+)- r_i) +(\chi^+-\pi^-) \overline{a}_i \\&~~- U_i(d^{\pi^-}_i)+ \pi^- (d_i^{\pi^-}-r_i).
    \end{align*}
    Since $d^{\psi^\ast}_i(\chi^+) < d^{\pi^-}_i$, we have
    \begin{align*}
    \pi^- (d^{\psi^\ast}_i(\chi^+)-d^{\pi^-}_i) &\geq U_i(d^{\psi^\ast}_i(\chi^+)) - U_i(d^{\pi^-}_i)\\&\geq \chi^+ (d^{\psi^\ast}_i(\chi^+)-d^{\pi^-}_i),
    \end{align*}
    and using the lower bound, we get
    $$\Delta S_i \geq (\chi^+ - \pi^-) (r_i - d^{\pi^-}_i + \overline{a}_i).$$
    We know that $\chi^+ \geq \pi^-$, so we only need to show that $r_i - d^{\pi^-}_i + \overline{a}_i\geq 0$.
    Note that by adding $\overline{a}_i$ to $d^{\pi^-}_i \leq r_i \leq d^{\pi^-}_i - \underline{z}_i$ (from case 4) we get
    $$0 \leq \overline{a}_i < r_i - d^{\pi^-}_i + \overline{a}_i < \overline{a}_i - \underline{z}_i,$$
    which proves that $\Delta S_i \geq (\chi^+ - \pi^-) (r_i - d^{\pi^-}_i + \overline{a}_i)\geq 0$.
  \item [(Case 5: $r_\Nc \leq \sigma_1, r_i \geq \Delta_4^i$)]  The surplus difference is
  \begin{align*}
   \Delta S_i&=U_i(d^{\psi^\ast}_i(\chi^+)) - \chi^+ ( d^{\psi^\ast}_i(\chi^+)- r_i) +(\chi^+-\pi^+) \overline{a}_i \\&~~- U_i(d^{\mu^-}_i)+ \pi^- \underline{z}_i.
    \end{align*}
    Since $d^{\psi^\ast}_i(\chi^+) < d^{\mu^-}_i$, we have
    \begin{align*}
    \mu^- (d^{\psi^\ast}_i(\chi^+)-d^{\mu^-}_i) &\geq U_i(d^{\psi^\ast}_i(\chi^+)) - U_i(d^{\mu^-}_i)\\&\geq \chi^+ (d^{\psi^\ast}_i(\chi^+)-d^{\mu^-}_i),
    \end{align*}
     and using the lower bound and $\underline{z}_i = d^{\mu^-}_i - r_i$, we get
    $$\Delta S_i \geq (\chi^+ - \mu^-) (r_i - d^{\mu^-}_i) +(\chi^+ - \pi^+) \overline{a}_i\geq 0,$$
    because $r_i \geq d^{\mu^-}_i$.
  \item [(Case 6: $r_\Nc \geq \sigma_4, r_i \leq \Delta_1^i$)] The surplus difference is
  \begin{align*}
   \Delta S_i&=U_i(d^{\psi^\ast}_i(\chi^-)) - \chi^- ( d^{\psi^\ast}_i(\chi^-)- r_i) +(\chi^--\pi^-) \underline{a}_i \\&~~- U_i(d^{\mu^+}_i)+ \pi^+ \overline{z}_i.
    \end{align*}
    Since $d^{\psi^\ast}_i(\chi^-) > d^{\mu^+}_i$, we have
     \begin{align*}
     \mu^+ (d^{\psi^\ast}_i(\chi^-)-d^{\mu^+}_i) &\geq U_i(d^{\psi^\ast}_i(\chi^-)) - U_i(d^{\mu^+}_i)\\&\geq \chi^- (d^{\psi^\ast}_i(\chi^-)-d^{\mu^+}_i),
     \end{align*}
    and using the lower bound and $\overline{z}_i= d^{\mu^+}_i - r_i$, we get
     $$\Delta S_i \geq (\chi^- - \pi^+) (r_i - d^{\mu^+}_i) +(\chi^- - \pi^-) \underline{a}_i\geq 0$$
     because $\chi^-\leq \pi^- \leq  \pi^+$, $r_i \leq d^{\mu^+}_i$ and $\underline{a}_i \leq 0$.
  \item [(Case 7: $r_\Nc \geq \sigma_4, \Delta_1^i < r_i < \Delta_2^i$)] The surplus difference is
  \begin{align*}
   \Delta S_i&=U_i(d^{\psi^\ast}_i(\chi^-)) - \chi^- ( d^{\psi^\ast}_i(\chi^-)- r_i) +(\chi^--\pi^-) \underline{a}_i \\&~~- U_i(d^{\pi^+}_i)+ \pi^+ (d_i^{\pi^+}-r_i).
    \end{align*}
     Since $d^{\psi^\ast}_i(\chi^-) > d^{\pi^+}_i$, we have
    \begin{align*}
    \pi^+ (d^{\psi^\ast}_i(\chi^-)-d^{\pi^+}_i) &\geq U_i(d^{\psi^\ast}_i(\chi^-)) - U_i(d^{\pi^+}_i)\\&\geq \chi^- (d^{\psi^\ast}_i(\chi^-)-d^{\pi^+}_i),
    \end{align*}
    and using the lower bound, we get
    $$\Delta S_i \geq (\chi^- - \pi^+) (r_i - d^{\pi^+}_i) +(\chi^- - \pi^-) \underline{a}_i\geq 0$$
    because $\chi^-\leq \pi^- \leq  \pi^+$ and $r_i \leq d^{\pi^+}_i$.
  \item [(Case 8: $r_\Nc \geq \sigma_4, \Delta_2^i \leq r_i \leq \Delta_3^i$)] The surplus difference is
  \begin{align*}
   \Delta S_i&=U_i(d^{\psi^\ast}_i(\chi^-)) - \chi^- ( d^{\psi^\ast}_i(\chi^-)- r_i) +(\chi^--\pi^-) \underline{a}_i\\&~~- U_i(d^{\mu^o}_i).
    \end{align*}
     Since $d^{\psi^\ast}_i(\chi^-) > d^{\mu^o}_i$, we have
     \begin{align*}
     \mu^o (d^{\psi^\ast}_i(\chi^-)-d^{\mu^o}_i) &\geq U_i(d^{\psi^\ast}_i(\chi^-)) - U_i(d^{\mu^o}_i)\\&\geq \chi^- (d^{\psi^\ast}_i(\chi^-)-d^{\mu^o}_i),
     \end{align*}
     and using the lower bound and $d^{\mu^o}_i = r_i$, we get
     $$(\chi^- - \pi^-) \underline{a}_i\geq 0.$$
  \item [(Case 9: $r_\Nc \geq \sigma_4, \Delta_3^i < r_i < \Delta_4^i$)] The surplus difference is
  \begin{align*}
   \Delta S_i&=U_i(d^{\psi^\ast}_i(\chi^-)) - \chi^- ( d^{\psi^\ast}_i(\chi^-)- r_i) +(\chi^--\pi^-) \underline{a}_i\\&~~- U_i(d^{\pi^-}_i)+ \pi^- (d^{\pi^-}_i - r_i).
    \end{align*}
     Since $d^{\psi^\ast}_i(\chi^-) > d^{\pi^-}_i$, we have
     \begin{align*}
     \pi^- (d^{\psi^\ast}_i(\chi^-)-d^{\pi^-}_i) &\geq U_i(d^{\psi^\ast}_i(\chi^-)) - U_i(d^{\pi^-}_i)\\&\geq \chi^- (d^{\psi^\ast}_i(\chi^-)-d^{\pi^-}_i),
     \end{align*}
     and using the lower bound we get
     $$\Delta S_i \geq(\chi^- - \pi^-) (r_i - d^{\pi^-}_i+\underline{a}_i).$$
     Note that $\chi^- -\pi^- \leq 0$ and by adding $\overline{a}_i$ to $\Delta_3^i < r_i < \Delta_4^i$ we get 
     $$\underline{a}_i < r_i- d^{\pi^-}_i+\underline{a}_i < -\underline{z}_i+\underline{a}_i \leq 0$$
     which proves that $\Delta S_i\geq 0$.
  \item [(Case 10: $r_\Nc \geq \sigma_4, r_i \geq \Delta_4^i$)] The surplus difference is
  \begin{align*}
   \Delta S_i&=U_i(d^{\psi^\ast}_i(\chi^-)) - \chi^- ( d^{\psi^\ast}_i(\chi^-)- r_i) +(\chi^--\pi^-) \underline{a}_i\\&- U_i(d^{\mu^-}_i)+ \pi^- \underline{z}_i.
    \end{align*}
Here we have three sub-cases i) $d^{\psi^\ast}_i(\chi^-)> d^{\mu^-}_i$, ii) $d^{\psi^\ast}_i(\chi^-)< d^{\mu^-}_i$, and iii) $d^{\psi^\ast}_i(\chi^-)= d^{\mu^-}_i$. Under sub-case (i), we can write
    \begin{align*}
    \mu^- (d^{\psi^\ast}_i(\chi^-)-d^{\mu^-}_i) &\geq U_i(d^{\psi^\ast}_i(\chi^-)) - U_i(d^{\mu^-}_i)\\&\geq \chi^- (d^{\psi^\ast}_i(\chi^-)-d^{\mu^-}_i),
    \end{align*}
    Using the lower bound, we get
    $$\Delta S_i \geq (\chi^--\pi^-)(\underline{a}_i-\underline{z}_i)\geq 0,$$
    where we used $\underline{z}_i= d^{\mu^-}_i - r_i$ when $r_i \geq \Delta_4^i$.\\
    Under sub-case (ii), we have
    \begin{align*}
    \chi^- (d^{\mu^-}_i-d^{\psi^\ast}_i(\chi^-)) &\geq  U_i(d^{\mu^-}_i) - U_i(d^{\psi^\ast}_i(\chi^-))\\&\geq \mu^- (d^{\mu^-}_i-d^{\psi^\ast}_i(\chi^-)).
    \end{align*}
    By multiplying the inequalities above by $-1$, we get the inequality in sub-case (i) and the proof follows directly.\\
    Under sub-case (iii), we have $\underline{z}_i = d^{\mu^-}_i - r_i = d^{\psi^\ast}_i(\chi^-) - r_i$, hence
    $$\Delta S_i \geq (\chi^- - \pi^-)(a_i - \underline{z}_i)\geq 0.$$
  \item [(Case 11: $\sigma_1 < r_\Nc < \sigma_2, r_i \leq \Delta_1^i$)] The surplus difference is
  \begin{align*}
   \Delta S_i&=U_i(d^{\psi^\ast}_i(\pi^+)) - \pi^+ ( d^{\psi^\ast}_i(\pi^+)- r_i)- U_i(d^{\mu^+}_i)+ \pi^+ \overline{z}_i.
    \end{align*}
    Since $d^{\psi^\ast}_i(\pi^+) > d^{\mu^+}_i$, we have
    \begin{align*}
    \mu^+ (d^{\psi^\ast}_i(\pi^+)-d^{\mu^+}_i) &\geq  U_i(d^{\psi^\ast}_i(\pi^+)) - U_i(d^{\mu^+}_i)\\&\geq \pi^+ (d^{\psi^\ast}_i(\pi^+)-d^{\mu^+}_i),
    \end{align*}
    and using the lower bound and $\overline{z}_i=d^{\mu^+}_i-r_i$, we get $\Delta S_i\geq 0$.
  \item [(Case 12: $\sigma_1 < r_\Nc < \sigma_2, r_i \geq \Delta_4^i$)] The surplus difference is
  \begin{align*}
   \Delta S_i&=U_i(d^{\psi^\ast}_i(\pi^+)) - \pi^+ ( d^{\psi^\ast}_i(\pi^+)- r_i)- U_i(d^{\mu^-}_i)+ \pi^- \underline{z}_i.
    \end{align*}
    Since $d^{\psi^\ast}_i(\pi^+) < d^{\mu^-}_i$, we have
    \begin{align*}
    \pi^+ (d^{\psi^\ast}_i(\pi^+)-d^{\mu^-}_i) &\leq  U_i(d^{\psi^\ast}_i(\pi^+)) - U_i(d^{\mu^-}_i)\\&\leq \mu^- (d^{\psi^\ast}_i(\pi^+)-d^{\mu^-}_i),
    \end{align*}
    and using the lower bound and $\underline{z}_i=d^{\mu^-}_i-r_i$, we get $\Delta S_i \geq (\pi^--\pi^+)\underline{z}_i\geq 0$.
  \item [(Case 13: $\sigma_2 \leq r_\Nc\leq \sigma_3, r_i \leq \Delta_1^i$)] The surplus difference is
  \begin{align*}
   \Delta S_i&=U_i(d^{\psi^\ast}_i(\chi^z)) - \chi^z ( d^{\psi^\ast}_i(\chi^z)- r_i)- U_i(d^{\mu^+}_i)+ \pi^+ \overline{z}_i.
    \end{align*}
    Since $d^{\psi^\ast}_i(\chi^z) > d^{\mu^+}_i$, we have
    \begin{align*}
    \mu^+ (d^{\psi^\ast}_i(\chi^z)-d^{\mu^+}_i) &\geq  U_i(d^{\psi^\ast}_i(\chi^z)) - U_i(d^{\mu^+}_i)\\&\geq \chi^z (d^{\psi^\ast}_i(\chi^z)-d^{\mu^+}_i),
    \end{align*}
    and using the lower bound and $\overline{z}_i=d^{\mu^+}_i-r_i$, we get $\Delta S_i \geq (\pi^+ - \chi^z)\overline{z}_i\geq 0$.
  \item [(Case 14: $\sigma_2 \leq r_\Nc\leq \sigma_3, r_i \geq \Delta_4^i$)] The surplus difference is
  \begin{align*}
   \Delta S_i&=U_i(d^{\psi^\ast}_i(\chi^z)) - \chi^z ( d^{\psi^\ast}_i(\chi^z)- r_i)- U_i(d^{\mu^-}_i)+ \pi^- \underline{z}_i.
    \end{align*}
    Since $d^{\psi^\ast}_i(\chi^z) < d^{\mu^-}_i$, we have
    \begin{align*}
    \chi^z (d^{\psi^\ast}_i(\chi^z)-d^{\mu^-}_i) &\leq  U_i(d^{\psi^\ast}_i(\chi^z)) - U_i(d^{\mu^-}_i)\\&\leq \mu^- (d^{\psi^\ast}_i(\chi^z)-d^{\mu^-}_i),
    \end{align*}
    and using the lower bound and $\underline{z}_i=d^{\mu^-}_i-r_i$, we get $\Delta S_i \geq (\pi^--\chi^z)\underline{z}_i\geq 0$.
   \item [(Case 15: $\sigma_3 < r_\Nc < \sigma_4, r_i \leq \Delta_1^i$)] The surplus difference is
  \begin{align*}
   \Delta S_i&=U_i(d^{\psi^\ast}_i(\pi^-)) - \pi^- ( d^{\psi^\ast}_i(\pi^-)- r_i)- U_i(d^{\mu^+}_i)+ \pi^+ \overline{z}_i.
    \end{align*}
    Since $d^{\psi^\ast}_i(\pi^-) > d^{\mu^+}_i$, we have
    \begin{align*}
    \mu^+ (d^{\psi^\ast}_i(\pi^-)-d^{\mu^+}_i) &\geq  U_i(d^{\psi^\ast}_i(\pi^-)) - U_i(d^{\mu^+}_i)\\&\geq \pi^- (d^{\psi^\ast}_i(\pi^-)-d^{\mu^+}_i),
    \end{align*}
    and using the lower bound and $\overline{z}_i=d^{\mu^+}_i-r_i$, we get $\Delta S_i \geq (\pi^+-\pi^-)\underline{z}_i\geq 0$.
   \item [(Case 16: $\sigma_3 < r_\Nc< \sigma_4, r_i \geq \Delta_4^i$)] The surplus difference is
  \begin{align*}
   \Delta S_i&=U_i(d^{\psi^\ast}_i(\pi^-)) - \pi^- ( d^{\psi^\ast}_i(\pi^-)- r_i)- U_i(d^{\mu^-}_i)+ \pi^- \underline{z}_i.
    \end{align*}
    Since $d^{\psi^\ast}_i(\pi^-) < d^{\mu^-}_i$, we have
    \begin{align*}
    \pi^- (d^{\psi^\ast}_i(\pi^-)-d^{\mu^-}_i) &\leq  U_i(d^{\psi^\ast}_i(\pi^-)) - U_i(d^{\mu^-}_i)\\&\leq \mu^- (d^{\psi^\ast}_i(\pi^-)-d^{\mu^-}_i),
    \end{align*}
    and using the lower bound and $\underline{z}_i=d^{\mu^-}_i-r_i$, we get $\Delta S_i\geq 0$.
\end{description}

%% file: appendixB_v2/WelfareLem_v2.tex
\begin{lemma}[Community maximum welfare]\label{lem:CentralizedWelfare}
   The maximum community welfare under centralized resource scheduling, given by, 
   \begin{align}
W^{\ast,\mbox{\tiny NEM}}_\Nc:= \underset{(d_i,\ldots,d_N)}{\rm maximize}&~~  \mathbb{E}\big[\sum_{i\in \Nc} U_i(d_i)-P^{\mbox{\tiny NEM}}(z_\Nc)\big]\nn \\ \text{subject to} &~~ z_\Nc = \sum_{i\in \Nc} \big( d_{i}-r_i\big) \nn\\&\hspace{0.4cm} \underline{z}_{\Nc} \leq z_{\Nc} \leq \overline{z}_{\Nc} \nn\\&\hspace{0.4cm}0< d_i < \overline{d}_i,~ \forall i\in \Nc.\nn
\end{align}
   is a monotonically increasing function of $r_{\Nc}$, given by
   \begin{align}
       &W^{\ast,\mbox{\tiny NEM}}_\Nc(r_{\Nc}) = \nn\\&\mbbE\begin{cases}
           \sum_{i\in \Nc} U_i(d_i^{\chi^+}(r_{\Nc})) - \pi^+ \overline{z}_\Nc,&\hspace{-.75em} r_{\Nc} \leq \sigma_1\\
           \sum_{i\in \Nc} U_i(d_i^{\pi^+}) - \pi^+ (\sum_{i\in \Nc}  d_i^{\pi^+}- r_{\Nc}),&\hspace{-.75em}  r_{\Nc} \in (\sigma_1,\sigma_2)\\
           \sum_{i\in \Nc} U_i(d_i^{\chi^z}(r_{\Nc})),&\hspace{-.75em}  r_{\Nc} \in [\sigma_2,\sigma_3]\\
            \sum_{i\in \Nc} U_i(d_i^{\pi^-}) - \pi^- (\sum_{i\in \Nc}  d_i^{\pi^-}- r_{\Nc}),&\hspace{-.75em}  r_{\Nc} \in (\sigma_3,\sigma_4)\\
            \sum_{i\in \Nc} U_i(d_i^{\chi^-}(r_{\Nc})) - \pi^- \underline{z}_\Nc,&\hspace{-.75em}  r_{\Nc} \geq \sigma_4,\label{eq:MaxCentralW}
       \end{cases}
   \end{align}
   where the thresholds $\sigma_1-\sigma_4$ are computed as in the pricing policy and all consumptions are given by $ d^x_{i} := [f_{i}(x)]^{\overline{d}_{i}}_{0}$.
\end{lemma}

\subsection*{Proof of Lemma \ref{lem:CentralizedWelfare}}
Given the assumptions $\underline{z}_i\leq  \overline{d}_{i} - r_i$, and $\sum_{i\in \Nc} \overline{z}_i \leq \overline{z}_\Nc,~~ \sum_{i\in \Nc} \underline{z}_i \geq \underline{z}_\Nc$, the proof of the maximum welfare of the community under centralized control follows the proof from Lemma \ref{lem:BenchmarkSur}, as it is a generalization of the benchmark customer problem by adding the dimension of community members. Therefore, the optimal aggregate consumption $d^{\ast}_\Nc$ and maximum welfare $W^{\ast,\mbox{\tiny NEM}}_\Nc$ will be functions of $r_{\Nc}:= \sum_{i\in \Nc} r_i$ rather than $r_i$. The optimal consumption of every member $i\in \Nc$ under centralized operation is, therefore,
\begin{equation*}
       d^{\ast}_i(r_{\Nc}) = \begin{cases}
           d_i^{\chi^+}(r_{\Nc}) &,  r_{\Nc} \leq \sigma_1\\
          d_i^{\pi^+}&,  r_{\Nc} \in (\sigma_1,\sigma_2)\\
           d_i^{\chi^z}(r_{\Nc})&,  r_{\Nc} \in [\sigma_2,\sigma_3]\\
            d_i^{\pi^-}&,  r_{\Nc} \in (\sigma_3,\sigma_4)\\
            d_i^{\chi^-}(r_{\Nc})&,  r_{\Nc} \geq \sigma_4,
       \end{cases}
   \end{equation*}
   which yields the aggregate net-consumption
   \begin{align*}
       z^{\ast}_{\Nc}(r_{\Nc}) = \begin{cases}
           \overline{z}_\Nc &,  r_{\Nc} \leq \sigma_1\\
           \sum_{i\in \Nc} d_i^{\pi^+} - r_{\Nc}&,  r_{\Nc} \in (\sigma_1,\sigma_2)\\
           0&,  r_{\Nc} \in [\sigma_2,\sigma_3]\\
             \sum_{i\in \Nc} d_i^{\pi^-} - r_{\Nc}&,  r_{\Nc} \in (\sigma_3,\sigma_4)\\
            \underline{z}_\Nc&,  r_{\Nc} \geq \sigma_4.
       \end{cases}
   \end{align*}
The maximum welfare can be directly computed by plugging $d^{\ast}_i(r_{\Nc})$ and $z^{\ast}_{\Nc}(r_{\Nc})$ into $W^{\ast,\mbox{\tiny NEM}}_\Nc (r_\Nc)= \sum_{i\in \Nc} U_i(d^{\ast}_i(r_{\Nc})) - P^{\mbox{\tiny NEM}}_\Nc(z^{\ast}_{\Nc}(r_{\Nc}))$, which yield (\ref{eq:MaxCentralW}). Note that the expectation is irrelevant here because the solution optimizes for every possible realization, hence it optimizes the expected value too.\hfill$\Box$

%% file: appendixB_v2/Equilibrium_v2.tex
The stochastic bi-level optimization in $\S$\ref{subsec:BilevelOpt} can be compactly formulated as 
\begin{align}
 \underset{\{P_i(\cdot)\}_{i=1}^N, \{d_i\}_{i=1}^N}{\operatorname{maximize}}& \Bigg(W^{\chi_\psi} = \mbbE\Big[\sum_{i\in \Nc} U_i(d_i^\psi) - P^{\mbox{\tiny NEM}}(z_{\Nc})\Big]\Bigg)\nn\\
 \text{subject to}&~~~ \underline{z}_{\Nc} \leq z_{\Nc} = \sum_{i\in \Nc}  d_i^\psi- r_i \leq \overline{z}_{\Nc}\nn\\
 &~~~ \text{ for all } i=1, \ldots, N\nn\\
 &~~~ d^{\psi}_i := \underset{d_i}{\operatorname{argmax}}~~ U_i(d_i) - P_i(z_i)\nn\\
 &\hspace{2.8cm} 0 \leq d_i \leq \overline{d}_i,\nn
\end{align}
where we also removed the constraint $\{P_i(\cdot)\}_{i=1}^N \in \Ac$ from the upper level problem, hence we solve an upper bound that does not constrain the payment rule to the set of cost-causation conforming payment rules $\Ac$. Next, we reformulate the program above by (1) replacing the lower-level optimization problem by its KKT conditions (mathematical program with equilibrium constraints \cite{Luo&Pang&Ralph:96Cambridge}), and (2) using $P_i(z_i):= \pi \cdot z_i, \forall i \in \Nc$ from Axioms \ref{ax:equity}--\ref{ax:monotonicity}, as
\begin{align}
 \underset{\pi, \{\overline{\bm{\lambda}}\}_{i=1}^N,\{\underline{\bm{\lambda}}\}_{i=1}^N}{\operatorname{maximize}}& \Bigg(W^{\chi_\psi} = \mbbE\Big[\sum_{i\in \Nc} U_i(d_i^\psi) - P^{\mbox{\tiny NEM}}(z_{\Nc})\Big]\Bigg)\nn\\
 \text{subject to}&~~~ \underline{z}_{\Nc} \leq z_{\Nc} = \sum_{i\in \Nc}  d_{i}^\psi- r_i \leq \overline{z}_{\Nc},\nn\\
 &~~~ \text{ for all } i=1, \ldots, N\nn\\
 &~~~ L_{i}(d_{i}^\psi)-\pi+ \overline{\lambda}_{i} -  \underline{\lambda}_{i}=0 \nn\\
 &\hspace{1.6cm} \overline{d}_{i} - d_{i}^\psi \geq 0 \perp \overline{\lambda}_{i} \geq 0 \nn\\
 &\hspace{2.35cm} d_{i}^\psi \geq  0 \perp \underline{\lambda}_{i} \geq 0, \nn
\end{align}
where $x \perp y$ means that $x$ and $y$ are perpendicular. By implementing the complementarity conditions above into the constraint $L_{i}(d_{i}^\psi)-\pi+ \overline{\lambda}_{i} -  \underline{\lambda}_{i}=0$, we have
\begin{align}
 \underset{\pi}{\operatorname{maximize}}& \Bigg(W^{\chi_\psi} = \mbbE\Big[\sum_{i\in \Nc} U_i(d_i^\psi) - P^{\mbox{\tiny NEM}}(z_{\Nc})\Big]\Bigg)\nn\\
 \text{subject to}&~~~ \underline{z}_{\Nc} \leq z_{\Nc} = \sum_{i\in \Nc}  d_{i}^\psi- r_i \leq \overline{z}_{\Nc},\nn\\
 &~~~ \text{ for all } i=1, \ldots, N\nn\\
 &~~~ d_{i}^\psi= \max\{0,\min\{f_{i}(\pi),\overline{d}_{i}\} \}. \nn
\end{align}
 Now, note that if $\pi^\ast$ is found, we have equilibrium. The concave and non-differentiable objective above can be divided into the following three convex programs $\mathcal{P}^{\mbox{\tiny NEM},+}, \mathcal{P}^{\mbox{\tiny NEM},-}$, and $\mathcal{P}^{\mbox{\tiny NEM},o}$, which correspond to when $z_{\Nc} \geq 0, z_{\Nc} \leq 0$ and $z_{\Nc}=0$, respectively:
\begin{align*}
\begin{array}{lll}\mathcal{P}^{\mbox{\tiny NEM},+}: &\hspace{-0.2cm}  \underset{\pi}{\rm minimize}& \hspace{-0.25cm} \sum_{i\in \Nc} \left(\pi^+ (  d_{i}^\psi(\pi) - r_{i})-U_i(d_{i}^\psi(\pi))\right)  \\&\text{subject to} & \overline{z}_{\Nc} \geq \sum_{i\in \Nc} \left( d_{i}^\psi(\pi) - r_{i}\right) \geq 0.
\end{array}   \end{align*} 
\begin{align*}
\begin{array}{lll}\mathcal{P}^{\mbox{\tiny NEM},-}: &\hspace{-0.2cm}  \underset{\pi}{\rm minimize}&\hspace{-0.25cm}  \sum_{i\in \Nc} \left(\pi^- (  d_{i}^\psi(\pi) - r_{i})-U_i(d_{i}^\psi(\pi))\right)  \\&\text{subject to} & 0 \geq \sum_{i\in \Nc} \left( d_{i}^\psi(\pi) - r_{i}\right) \geq \underline{z}_{\Nc}.
\end{array}   \end{align*} 
\begin{align*}
\begin{array}{lll}\mathcal{P}^{\mbox{\tiny NEM},o}: &  \underset{\pi}{\rm minimize}&  - \sum_{i\in \Nc} U_i(d_{i}^\psi(\pi))  \\&\text{subject to} & \sum_{i\in \Nc}\left( d_{i}^\psi(\pi) - r_{i}\right) = 0. 
\end{array}   \end{align*} 
Starting with $\mathcal{P}^{\mbox{\tiny NEM},+}$, the Lagrangian $\Lc^+$ is given by
    \begin{align*}
        \Lc^+(\cdot) =& \sum_{i\in \Nc} \pi^+ ( d_{i}^\psi(\pi) - r_i) - \sum_{i\in \Nc}  U_{i}(d_{i}^\psi(\pi))\\& +  \overline{\gamma}^+ \big(\sum_{i\in \Nc} ( d_{i}^\psi(\pi)-r_i) -\overline{z}_{\Nc}\big)\\&- \underline{\gamma}^+ \sum_{i\in \Nc} ( d_{i}^\psi(\pi)-r_i)
    \end{align*}
here $\overline{\gamma}^+,\underline{\gamma}^+\geq 0$ are Lagrange multipliers for the import OE and non-negative aggregate net-consumption constraint. Given that $\mathcal{P}^{\mbox{\tiny NEM},o}$ covers the case when $\underline{\gamma}^+>0$, we can here set $\underline{\gamma}^+=0$. From the KKT conditions we have,
$$\frac{\partial \Lc^+(\cdot)}{\partial \pi}= \sum_{i\in \Nc}  \frac{\partial d_{i}^\psi(\pi)}{\partial \pi}\left(\pi^+ -  L_{i}(d_{i}^\psi(\pi^\ast))+ \overline{\gamma}^+\right)=0.$$
If $\overline{\gamma}^+=0$, then $\pi^+ = L_{i}(d_{i}^\psi(\pi^\ast))$, and because $f_{i} = L^{-1}_{i}$, it must be that $\pi^\ast= \pi^+$ when $r_{\Nc} > \sum_{i\in \Nc}  d_{i}^\psi(\pi^+)- \overline{z}_{\Nc}=:\sigma_1$.

On the other hand, if $\overline{\gamma}^+>0$, then $\pi^\ast = \pi^+ + \overline{\gamma}^+$ and the price must be so that the following equality holds
\begin{equation*}
    \overline{z}_{\Nc} + r_{\Nc} = \sum_{i\in \Nc}  d_{i}^\psi(\pi^+ + \overline{\gamma}^+),
\end{equation*}
which is the same equality to find $\chi^+(r_{\Nc})$ in (\ref{eq:MonotonicPrices1}), hence $\pi^\ast(r_{\Nc}) = \chi^+(r_{\Nc})\geq \pi^+$. Proposition \ref{corol:PriceOrder} shows that $\chi^+(r_\Nc)$ is non-negative, continuous and monotonically decreasing with $r_\Nc$, which implies that ${d}^{\chi^+}_{i} \in [0,d_{i}^{\pi^+}]$.

Following the same steps as in $\mathcal{P}^{\mbox{\tiny NEM},+}$, we can show that by solving $\mathcal{P}^{\mbox{\tiny NEM},-}$, we have $\pi^\ast(r_{\Nc}) = \pi^-$ when $r_{\Nc}<\sigma_4$ and $\underline{z}_{\Nc}<\sum_{i\in \Nc} ( d_{i}^\psi(\pi^-) - r_{i})<0$ and $\pi^\ast(r_{\Nc}) = \chi^-(r_{\Nc})$  when $r_{\Nc}\geq \sigma_4$ and $\sum_{i\in \Nc} ( d_{i}^\psi(\chi^-(r_{\Nc})) - r_{i}) = \underline{z}_{\Nc}$.

Under $\mathcal{P}^{\mbox{\tiny NEM},o}$, the Lagrangian $\Lc^o$ is given by
\begin{equation*}
        \Lc^o(\cdot) = - \sum_{i\in \Nc}  U_{i}(d_{i}^\psi(\pi)) +  \eta^o \sum_{i\in \Nc} ( d_{i}^\psi(\pi)-r_i),
    \end{equation*}
where $\eta^o \geq 0$ is the Lagrangian multiplier of the equality constraint. By KKT conditions, the optimal price is $\pi^\ast(r_{\Nc})=\eta^o(r_{\Nc})$, where $\eta^o(r_{\Nc})$ must be such that the equality constraint holds
\begin{equation*}
    \sum_{i\in \Nc}  d_{i}^\psi(\pi)= r_{\Nc},
\end{equation*}
which is the same equality to find $\chi^z(r_{\Nc})$ in (\ref{eq:MonotonicPrices2}), hence $\pi^\ast(r_{\Nc})= \chi^z(r_{\Nc})$. Now we will show that $\chi^z(r_{\Nc}) \in [\pi^-, \pi^+]$ when $r_{\Nc} \in [\sigma_2,\sigma_3]$, let
\begin{equation}\label{eq:Thm1F2}
    F_2(x):= \sum_{i\in \Nc}  d_{i}^\psi(x)-r_{\Nc},
\end{equation}
Note that $F_2(x)$ is continuous and monotonically decreasing. Because
$$F_2(\pi^+)\leq 0, ~~ F_2(\pi^-)\geq 0,$$
there must exists $\pi \in [\pi^-, \pi^+]$ such that $F_2(\pi)=0$, and from the continuity and monotonicity of $F_2$ in $r_{\Nc}$, and the monotonicity of the marginal utility function $L_i$, the price $\chi^z(r_{\Nc})$ is a continuous and monotonically decreasing function of $r_{\Nc}$. Hence, in summary $\pi^\ast(r_{\Nc})= \chi^z(r_{\Nc}) \in [\pi^-, \pi^+]$ when $r_{\Nc} \in [\sigma_2,\sigma_3]$.

The equilibrium price $\pi^\ast(r_{\Nc})$ is therefore as shown in (\ref{eq:PricingMechanism}) with the payment rule $P^{\chi^\ast}_i(z_i)= \pi^\ast(r_{\Nc}) \cdot z_i - A^\ast_i(r_{\Nc})$ that belongs to $\Ac$ as shown in Lemma \ref{lem:CostCausation}.

Lastly, to show that the equilibrium achieves the highest community welfare, we leverage Lemma \ref{lem:CentralizedWelfare}. Note that under the OEs-aware pricing, the member's optimal consumption policy
$$
d_{i}^{\psi^\ast}(\pi^\ast) = [f_{i}(\pi^\ast)]^{\overline{d}_{i}}_{0}, \forall i\in \Nc,
$$
results in the following surplus function
$$S^{\chi^\ast}_i(\psi^\ast(r_i),r_i) = U_i(d_{i}^{\psi^\ast}(\pi^\ast)) - P^{\chi^\ast}_i(z_{i}^{\psi^\ast}(\pi^\ast)),$$
where $z_{i}^{\psi^\ast}(\pi^\ast) =  d_{i}^{\psi^\ast}(\pi^\ast) - r_i$. By aggregating the surplus of members under their optimal consumption policy
\begin{align}
    &\hspace{-0.1cm}\sum_{i\in \Nc}\hspace{-0.08cm}S^{\chi^\ast}_i(\psi^\ast(r_i),r_i) \hspace{-0.1cm}=\hspace{-0.18cm} \sum_{i\in \Nc} \Big(U_i(d_{i}^{\psi^\ast}(\pi^\ast)) - P^{\chi^\ast}_i(z_{i}^{\psi^\ast}(\pi^\ast))\Big)\nn\\\stackrel{\text{(A)}}{=}& \hspace{-0.08cm}\sum_{i\in \Nc} \begin{cases} U_i(d_{i}^{\psi^\ast}(\chi^+)) -  \chi^+ z_{i}^{\psi^\ast}(\chi^+)&, r_\Nc\leq \sigma_1 \\ U_i(d_{i}^{\psi^\ast}(\pi^+)) -  \pi^+ z_{i}^{\psi^\ast}(\pi^+)&, r_\Nc\in (\sigma_1,\sigma_2) \\ U_i(d_{i}^{\psi^\ast}(\pi^z)) -  \pi^z z_{i}^{\psi^\ast}(\pi^z) &, r_\Nc\in[\sigma_2,\sigma_3] \\ U_i(d_{i}^{\psi^\ast}(\pi^-)) -  \pi^- z_{i}^{\psi^\ast}(\pi^-)&, r_\Nc\in (\sigma_3,\sigma_4)\\ U_i(d_{i}^{\psi^\ast}(\chi^-)) -  \chi^- z_{i}^{\psi^\ast}(\chi^-)&, r_\Nc \geq \sigma_4
    \end{cases}\nn\\&+ \begin{cases}
(\chi^+(r_{\Nc})-\pi^+) \overline{z}_\Nc & ,r_{\Nc}\leq \sigma_1 \\ 
(\chi^-(r_{\Nc})-\pi^-) \underline{z}_\Nc &, r_{\Nc}\geq \sigma_4\\
0&, \text{otherwise}.
\end{cases}\nn\\
    \stackrel{\text{(\ref{eq:AggOptz})}}{=}&  \begin{cases} \sum_{i\in \Nc} U_i(d_{i}^{\psi^\ast}(\chi^+)) -  \pi^+ \overline{z}_{\Nc}&, r_\Nc \leq \sigma_1 \\ \sum_{i\in \Nc} U_i(d_{i}^{\psi^\ast}(\pi^+)) -  \pi^+ z_{\Nc}^{\psi^\ast}(\pi^+)&, r_\Nc\in (\sigma_1,\sigma_2) \\ \sum_{i\in \Nc} U_i(d_{i}^{\psi^\ast}(\pi^z))  &, r_\Nc\in[\sigma_2,\sigma_3] \\ \sum_{i\in \Nc} U_i(d_{i}^{\psi^\ast}(\pi^-)) -  \pi^- z_{\Nc}^{\psi^\ast}(\pi^-)&,
 r_\Nc\in (\sigma_3,\sigma_4)\\ \sum_{i\in \Nc} U_i(d_{i}^{\psi^\ast}(\chi^-)) -  \pi^- \underline{z}_{\Nc} &, r_\Nc \geq \sigma_4 \end{cases}
 \nn\\=:& W^{\chi^\ast,\psi^\ast}, \nn
\end{align}
where (A) is done by leveraging D-NEM threshold structure, and the expectation is dropped because the solution is for every possible realization, therefore it includes the expected value too. From Lemma \ref{lem:CentralizedWelfare}, we can see that $W^{\chi^\ast,\psi^\ast} = W^{\ast,\mbox{\tiny NEM}}_\Nc(r_{\Nc})$, and that $d^{\psi^\ast}_i = d^{\sharp}_i, z^{\psi^\ast}_i = z^{\sharp}_i$, for all $i \in \Nc$, which proves that the community welfare under the equilibrium solution $W^{\chi^\ast,\psi^\ast}$ is the maximum community welfare under centralized operation $W^{\ast,\mbox{\tiny NEM}}_\Nc$. 
\hfill$\Box$

%% file: appendixB_v2/PriceOrderLem.tex
The proof uses the monotonicity of the inverse marginal utility function and the assumption $\pi^- \leq \pi^+$. For brevity, we show the monotonicity and order of $\chi^+(r_\Nc)$, and leave the proof of $\chi^z(r_\Nc)$ and $\chi^-(r_\Nc)$ to the reader.

Let 
\begin{equation*}
    F(\mu_1):= \sum_{i \in \Nc} [{f}_{i}(\mu_1)]^{\overline{{d}}_{i}}_{0} - r_{\Nc}- \overline{z}_\Nc,
\end{equation*}
which is continuous and monotonically decreasing with $r_\Nc$, and let $\mu_1^{\mbox{\tiny max}}\geq 0$ be the maximum price such that $[{f}_{i}(\mu_1^{\mbox{\tiny max}})]^{\overline{{d}}_{i}}_{0}=0, \forall i\in \Nc$. Because when $r_\Nc \leq \sigma_1$, we have $F(\pi^+)\geq 0$ and $F(\mu_{\mbox{\tiny max}})\leq 0$, there must exist a $\chi^+(r_{\Nc}) \in [\pi^+,\mu_{\mbox{\tiny max}}]$ such that $F(\chi^+(r_{\Nc}))=0$. From the continuity and monotonicity of $F$ in $r_{\Nc}$, and the monotonicity of the inverse marginal utility function $f_i$, the price $\chi^+(r_{\Nc})$ is a continuous and monotonically decreasing function of $r_{\Nc}$. 
\hfill$\Box$

%% file: appendixB_v2/VoCPve.tex
The proof follows directly form the proof of conforming with Axiom \ref{ax:rationality} in Lemma \ref{lem:CostCausation}, when the conditions $\pi^-<\pi^+$,  $\sum_{i\in \Nc} \overline{z}_i < \overline{z}_\Nc$ and $\sum_{i\in \Nc} \underline{z}_i > \underline{z}_\Nc$ are met. \hfill$\Box$

%% file: appendixB_v2/CompStatics.tex
Recall the surplus of the community member $S^{\chi^\ast}_i(\psi^\ast(r_i),r_i)$ and the maximum surplus of the benchmark in (\ref{eq:SurplusBench}). For every $i\in \Nc$, to prove the monotonicity of $\mbox{VoC}_i^{\chi^\ast,\mbox{\tiny NEM}}$ with each exogenous parameter we take the derivative with respect to that parameter.

1) Monotonicity with ($\overline{z}_{\Nc},\underline{z}_{\Nc}$):
Deriving $\mbox{VoC}_i^{\chi^\ast,\mbox{\tiny NEM}}$ with respect to $\overline{z}_{\Nc}$, we have
$$
\frac{\partial \mbox{VoC}_i^{\chi^\ast,\mbox{\tiny NEM}}}{\partial \overline{z}_{\Nc}}= \mbbE \bigg[\begin{cases}
(\chi^+(r_\Nc)-\pi^+)/N &, r_\Nc \leq \sigma_1 \\ 
0 &, \text{otherwise} 
\end{cases}\bigg]\geq 0,
$$
because $\chi^+(r_\Nc) \geq \pi^+$.

On the other hand, deriving $\mbox{VoC}_i^{\chi^\ast,\mbox{\tiny NEM}}$ w.r.t. $\underline{z}_{\Nc}$, we have
$$
\frac{\partial \mbox{VoC}_i^{\chi^\ast,\mbox{\tiny NEM}}}{\partial \underline{z}_{\Nc}}= \mbbE \bigg[\begin{cases}
(\chi^-(r_\Nc)-\pi^-)/N &, r_\Nc \geq \sigma_4 \\ 
0 &, \text{otherwise} 
\end{cases}\bigg]\leq 0,
$$
because $\chi^-(r_\Nc) \leq \pi^-$.

2) Monotonicity with ($\pi^+$):
\begin{align*}
\frac{\partial \mbox{VoC}_i^{\chi^\ast,\mbox{\tiny NEM}}}{\partial \pi^+}=& \mbbE \bigg[\begin{cases}
-\overline{a}_i &, r_\Nc \leq \sigma_1 \ \\ 
r_i - d_i^{\psi^\ast}(\pi^+) &,  r_\Nc \in (\sigma_1,\sigma_2)\\
0 &, r_\Nc \geq \sigma_2 
\end{cases}\\& + \begin{cases}
\overline{z}_i &, r_i \leq \Delta_1^i \ \\ 
 d_i^{\psi^\ast}(\pi^+)-r_i &,  r_i \in (\Delta_1^i,\Delta_2^i)\\
0 &, r_i \geq \Delta_2^i
\end{cases}\bigg],
\end{align*}
where we used $L_i(d_i^{\psi^\ast}(\pi^+))=\pi^+$, $d_i^{\pi^+}=d_i{^\psi}(\pi^+)$ and $\overline{a}_i$ is as in (\ref{eq:ais}). \\
\noindent 2.a) \underline{If $r_\Nc \leq \sigma_1$:} then we have 
\begin{align*}
\frac{\partial \mbox{VoC}_i^{\chi^\ast,\mbox{\tiny NEM}}}{\partial \pi^+}&=\mbbE \bigg[\begin{cases}
-(\overline{z}_\Nc-\sum_{i\in \Nc}\overline{z}_i)/N &, r_i \leq \Delta_1^i \\ 
 d_i^{\psi^\ast}(\pi^+)-r_i-\overline{a}_i &,  r_i \in (\Delta_1^i,\Delta_2^i)\\
-\overline{a}_i &, r_i \geq \Delta_2^i
\end{cases}\bigg]\\&\leq 0
\end{align*}
where in the first and third pieces above, we used the assumption, $\overline{z}_\Nc-\sum_{i\in \Nc}\overline{z}_i\geq 0$, and in the second piece we have from the thresholds
\begin{align*}
d_i^{\psi^\ast}(\pi^+)-r_i-\overline{a}_i\leq \Delta^i_1:= d_i^{\psi^\ast}(\pi^+) - \overline{z}_i - r_i < 0,
\end{align*}
because $\Delta^i_1 < r_i$.

\noindent 2.b) \underline{If $r_\Nc \in (\sigma_1,\sigma_2)$:} then we have 
\begin{align*}
\frac{\partial \mbox{VoC}_i^{\chi^\ast,\mbox{\tiny NEM}}}{\partial \pi^+}&=\mbbE \bigg[\begin{cases}
r_i-d_i^{\psi^\ast}(\pi^+)+\overline{z}_i &, r_i \leq \Delta_1^i \\ 
 0 &,  r_i \in (\Delta_1^i,\Delta_2^i)\\
r_i-d_i^{\psi^\ast}(\pi^+) &, r_i \geq \Delta_2^i
\end{cases}\bigg].
\end{align*}
When $r_i \leq \Delta_1^i$, we have $\frac{\partial \mbox{VoC}_i^{\chi^\ast,\mbox{\tiny NEM}}}{\partial \pi^+}\leq 0$ because $r_i \leq d_i^{\psi^\ast}(\pi^+) - \overline{z}_i:=\Delta_1^i$. When $r_i \geq \Delta_2^i\geq 0$, we have $\frac{\partial \mbox{VoC}_i^{\chi^\ast,\mbox{\tiny NEM}}}{\partial \pi^+}\geq 0 $ because $r_i \geq d_i^{\psi^\ast}(\pi^+):=\Delta_2^i$.

\noindent 2.c) \underline{If $r_\Nc \geq \sigma_2$:} then we have
\begin{align*}
\frac{\partial \mbox{VoC}_i^{\chi^\ast,\mbox{\tiny NEM}}}{\partial \pi^+}&=\mbbE \bigg[\begin{cases}
\overline{z}_i &, r_i \leq \Delta_1^i \\ 
 d_i^{\psi^\ast}(\pi^+)-r_i &,  r_i \in (\Delta_1^i,\Delta_2^i)\\
0 &, r_i \geq \Delta_2^i
\end{cases}\bigg]\\& \geq 0.
\end{align*}

3) Monotonicity with ($\pi^-$): \begin{align*}
\frac{\partial \mbox{VoC}_i^{\chi^\ast,\mbox{\tiny NEM}}}{\partial \pi^-}=& \mbbE \bigg[\begin{cases}
0 &, r_\Nc \leq \sigma_3 \ \\ 
r_i - d_i^{\psi^\ast}(\pi^-) &,  r_\Nc \in (\sigma_3,\sigma_4)\\
-\underline{a}_i &, r_\Nc \geq \sigma_4 
\end{cases}\\& + \begin{cases}
0 &, r_i \leq \Delta_3^i \ \\ 
 d_i^{\psi^\ast}(\pi^-)-r_i &,  r_i \in (\Delta_3^i,\Delta_4^i)\\
\underline{z}_i &, r_i \geq \Delta_4^i
\end{cases}\bigg],
\end{align*}
where we used $L_i(d_i^{\psi^\ast}(\pi^-))=\pi^-$, $d_i^{\pi^-}=d_i{^\psi}(\pi^-)$ and $\underline{a}_i$ is as in (\ref{eq:ais}). The monotonicity of each piece above can be easily shown by following the steps in the proof of the monotonicity with $\pi^+$.\hfill$\Box$

%% file: appendixB_v2/SurplusLem.tex
We compare the surplus difference between community members under aggregate-level OEs and member-level OEs frameworks, where Under the former, the customer's payment is based on the two-part pricing, whereas under the latter the customer's payment is one-part, as shown in $\S$\ref{sec:MemberLevelOEsPolicy}. The proof is complete if we show that
\begin{align*}
 &\Delta S^{\chi^\ast,\mbox{\tiny DNEM}^\ast}:= S^{\chi^\ast}_i(\psi^\ast(r_i),r_i)-S^{\mbox{\tiny DNEM}^\ast}_i(r_i)=\\&
    \begin{cases} U_i(d_{i}^{\psi^\ast}(\chi^+)) -  \chi^+ z_{i}^{\psi^\ast}(\chi^+) + (\chi^+-\pi^+)\overline{a}_i&\hspace{-0.3cm}, r_\Nc\leq \sigma_1 \\ U_i(d_{i}^{\psi^\ast}(\pi^+)) -  \pi^+ z_{i}^{\psi^\ast}(\pi^+)&\hspace{-0.3cm}, r_\Nc\in (\sigma_1,\sigma_2) \\ U_i(d_{i}^{\psi^\ast}(\pi^z)) -  \pi^z z_{i}^{\psi^\ast}(\pi^z) &\hspace{-0.3cm}, r_\Nc\in[\sigma_2,\sigma_3] \\ U_i(d_{i}^{\psi^\ast}(\pi^-)) -  \pi^- z_{i}^{\psi^\ast}(\pi^-)&\hspace{-0.3cm}, r_\Nc\in (\sigma_3,\sigma_4)\\ U_i(d_{i}^{\psi^\ast}(\chi^-)) -  \chi^- z_{i}^{\psi^\ast}(\chi^-) + (\chi^--\pi^-)\underline{a}_i&\hspace{-0.3cm}, r_\Nc \geq \sigma_4
    \end{cases}-\\& \begin{cases} 
U_i(d^{\mu^{\overline{z}}}_i(r_i)) - \pi^{\mbox{\tiny DNEM}}(\bm{r}) \overline{z}_i &, r_i < d^{\pi^{\mbox{\tiny DNEM}}}_i - \overline{z}_i\\ 
U_i(d^{\pi^{\mbox{\tiny DNEM}}}_i) - \pi^{\mbox{\tiny DNEM}}(\bm{r}) (d^{\pi^{\mbox{\tiny DNEM}}}_i - r_i) &\hspace{-0.3cm},  r_i-d^{\pi^{\mbox{\tiny DNEM}}}_i \in [ - \overline{z}_i, - \underline{z}_i]\\
U_i(d^{\mu^{\underline{z}}}_i(r_i)) - \pi^{\mbox{\tiny DNEM}}(\bm{r}) \underline{z}_i &, r_i > d^{\pi^{\mbox{\tiny DNEM}}}_i - \underline{z}_i,
\end{cases}
\end{align*}
is non-negative, where the prices $\mu^{\overline{z}}(r_i)$ and $\mu^{\underline{z}}(r_i)$ are the solutions of 
\begin{align*}
    [f_i(\mu_1)]_{0}^{\overline{d}_i} =  \overline{z}_i +r_i, ~~~
    [f_i(\mu_2)]_{0}^{\overline{d}_i} = \underline{z}_i+r_i,
\end{align*}
respectively. From the community pricing policy under member-level OEs in $\S$\ref{sec:MemberLevelOEsPolicy}, we have two main cases: 1) $r_{\Nc}< \Theta_1(\bm{r})$, 2) $r_{\Nc}> \Theta_2(\bm{r})$.

\noindent \underline{Case 1) $r_{\Nc}< \Theta_1(\bm{r}), r_{\Nc} < \sigma_2$:} Note that $\pi^{\mbox{\tiny DNEM}} = \pi^+$. We have two sub-cases for the aggregate-level OEs community: 1.a) $r_{\Nc}\leq \sigma_1$, and 1.b) $r_{\Nc} \in (\sigma_1,\sigma_2)$.

\noindent \underline{Case 1.a) $r_{\Nc}\leq \sigma_1$:} If $r_i \in [d^{\pi^+}_i - \overline{z}_i,d^{\pi^+}_i - \underline{z}_i]$ we simply have Case 2 in the proof of Axiom \ref{ax:rationality} in Lemma \ref{lem:CostCausation}. If $r_i > d^{\pi^+}_i - \underline{z}_i$, we know that $d^{\mu^{\underline{z}}}_i(r_i)> d_{i}^{\psi^\ast}(\chi^+)$, hence and using the concavity of the utility function, we have 
\begin{align*}
  \mu^{\underline{z}} (d^{\psi^\ast}_i(\chi^+)-d^{\mu^{\underline{z}}}_i(r_i)) &\geq U_i(d^{\psi^\ast}_i(\chi^+)) - U_i(d^{\mu^{\underline{z}}}_i(r_i))\\&\geq \chi^+ (d^{\psi^\ast}_i(\chi^+)-d^{\mu^{\underline{z}}}_i(r_i)),
  \end{align*}
and using the lower bound, we get
\begin{align*}\Delta S^{\chi^\ast,\mbox{\tiny DNEM}^\ast}&\geq (\chi^+ - \pi^+)\overline{a}_i + \chi^+ (r_i-d^{\mu^{\underline{z}}}_i(r_i)) + \pi^+ \underline{z}_i\\& = (\chi^+ - \pi^+)(\frac{\overline{z}_\Nc-\sum_{i\in \Nc}\overline{z}_i}{N})\geq 0,
\end{align*}
where we used $d^{\mu^{\underline{z}}}_i(r_i) = \underline{z}_i+r_i$. Lastly, if $r_i<d^{\pi^+}_i - \overline{z}_i$ then $d^{\mu^{\overline{z}}}_i(r_i) \lesseqgtr d^{\psi^\ast}_i(\chi^+)$. After checking both cases, one can see that $\Delta S^{\chi^\ast,\mbox{\tiny DNEM}^\ast} \geq 0$.

\noindent \underline{Case 1.b) $r_{\Nc} \in (\sigma_1,\sigma_2)$:} If $r_i \in [d^{\pi^+}_i - \overline{z}_i,d^{\pi^+}_i - \underline{z}_i]$ then $\Delta S^{\chi^\ast,\mbox{\tiny DNEM}^\ast} = 0$. If $r_i > d^{\pi^+}_i - \underline{z}_i$, we know that $d^{\mu^{\underline{z}}}_i(r_i)> d_{i}^{\psi^\ast}(\pi^+)$. Using the concave utility function property as above, and taking the lower bound, we get $\Delta S^{\chi^\ast,\mbox{\tiny DNEM}^\ast}\geq 0 $. Lastly, if $r_i<d^{\pi^+}_i - \overline{z}_i$ then $d^{\mu^{\overline{z}}}_i(r_i) < d^{\psi^\ast}_i(\pi^+)$, and using the concave utility function property as above, and taking the lower bound, we get $\Delta S^{\chi^\ast,\mbox{\tiny DNEM}^\ast}\geq 0 $.

\noindent \underline{Case 2) $r_{\Nc}> \Theta_2(\bm{r}), r_{\Nc} > \sigma_3$:} Note that $\pi^{\mbox{\tiny DNEM}} = \pi^-$. We have two sub-cases for the aggregate-level OEs community: 2.a) $r_{\Nc} \in (\sigma_3,\sigma_4)$, and 2.b) $r_{\Nc} \geq \sigma_4$.

\noindent \underline{Case 2.a) $r_{\Nc} \in (\sigma_3,\sigma_4)$:} If $r_i \in [d^{\pi^-}_i - \overline{z}_i,d^{\pi^-}_i - \underline{z}_i]$ we have $\Delta S^{\chi^\ast,\mbox{\tiny DNEM}^\ast}= 0 $. If $r_i > d^{\pi^-}_i - \underline{z}_i$, we know that $d^{\mu^{\underline{z}}}_i(r_i)> d_{i}^{\psi^\ast}(\pi^-)$. Using the concave utility function property as above, and taking the lower bound, we get $\Delta S^{\chi^\ast,\mbox{\tiny DNEM}^\ast}\geq 0 $. Lastly, if $r_i<d^{\pi^-}_i - \overline{z}_i$ then $d^{\mu^{\overline{z}}}_i(r_i) < d^{\psi^\ast}_i(\pi^-)$, and using the concave utility function property as above, and taking the lower bound, we get $\Delta S^{\chi^\ast,\mbox{\tiny DNEM}^\ast}\geq 0 $.

\noindent \underline{Case 2.b) $r_{\Nc} \geq \sigma_4$:} If $r_i \in [d^{\pi^-}_i - \overline{z}_i,d^{\pi^-}_i - \underline{z}_i]$ we simply have Case 9 in the proof of Axiom \ref{ax:rationality} in Lemma \ref{lem:CostCausation}. If $r_i < d^{\pi^-}_i - \overline{z}_i$, we know that $d^{\mu^{\overline{z}}}_i(r_i)< d_{i}^{\psi^\ast}(\chi^-)$, hence and using the concavity of the utility function, and taking the lower bound, we get
\begin{align*}\Delta S^{\chi^\ast,\mbox{\tiny DNEM}^\ast}\geq (\chi^- - \pi^-)(\frac{\underline{z}_\Nc-\sum_{i\in \Nc}\underline{z}_i}{N})\geq 0,
\end{align*}
where we used $d^{\mu^{\overline{z}}}_i(r_i) = \overline{z}_i+r_i$. Lastly, if $r_i>d^{\pi^-}_i - \underline{z}_i$ then $d^{\mu^{\underline{z}}}_i(r_i) \lesseqgtr d^{\psi^\ast}_i(\chi^-)$. After checking both cases, one can see that $\Delta S^{\chi^\ast,\mbox{\tiny DNEM}^\ast} \geq 0$. \hfill$\Box$

%\noindent \underline{Case 3) $r_{\Nc}\in [\Theta_1(\bm{r}),\Theta_2(\bm{r})],\pi^{\mbox{\tiny DNEM}} = \pi^z(\bm{r})$:}